\documentclass[journal]{IEEEtran}
\usepackage{hyperref}
\usepackage[all]{hypcap}        % needed
\usepackage{xcolor}
\usepackage[linesnumbered,ruled,vlined]{algorithm2e}
\usepackage{algorithmic}

\SetCommentSty{mycommfont}
\SetKwInput{KwInput}{Input}% Set the Input
\SetKwInput{KwInitialize}{Initialize}% set the Output
\SetKwInput{KwOutput}{Output} % set the Output
\usepackage{xcolor,soul,framed} %,caption
\usepackage{xspace}
\colorlet{shadecolor}{yellow}
\usepackage[pdftex]{graphicx}
\graphicspath{{../pdf/}{../jpeg/}}
\DeclareGraphicsExtensions{.pdf,.jpeg,.png}

\usepackage[cmex10]{amsmath}
\usepackage{array}
\usepackage{mdwmath}
\usepackage{mdwtab}
\usepackage{eqparbox}
\usepackage{url}
\usepackage{cite}

\usepackage{multirow}
\usepackage{amsmath,amssymb,amsfonts}
\usepackage{mathtools}
\usepackage{mathrsfs}
\begin{document}
\bstctlcite{IEEEexample:BSTcontrol}
    \title{Backhaul-Aware Intelligent Positioning of UAVs and Association of Terrestrial Base Stations for Fronthaul Connectivity}
  \author{Muhammad K.~Shehzad,~\IEEEmembership{Student Member,~IEEE,}
      Arsalan~Ahmad,~\IEEEmembership{Member,~IEEE}\\
      Syed Ali~Hassan,~\IEEEmembership{Senior Member,~IEEE,}
      and~Haejoon~Jung,~\IEEEmembership{Senior Member,~IEEE}

 % \thanks{Manuscript received July 30, 2020.}
  \thanks{M. K. Shehzad, A. Ahmad, and S. A. Hassan are with the School of Electrical Engineering \& Computer Science (SEECS), National University of Sciences \& Technology (NUST), Islamabad, Pakistan. (e-mails: \{mshehzad.msee17seecs, arsalan.ahmad, ali.hassan\}@seecs.edu.pk.)}
  \thanks{H. Jung is with the Department of Information and Telecommunication
Engineering, Incheon National University, Incheon 22012, South Korea. (e-mail: haejoonjung@inu.ac.kr).}
  % <-this % stops a space
  %\thanks{T. Reveyrand is with the XLIM Laboratory, UMR 7252, University of Limoges, 87060 Limoges, France (e-mail: tibault.reveyrand@xlim.fr).}%
}

\markboth{IEEE TRANSACTIONS ON NETWORK SCIENCE AND ENGINEERING, 2021%VOL.~xx, NO.~xx, January~2020
}{Muhammad \MakeLowercase{\textit{et al.}}: Backhaul-Aware Intelligent Positioning of UAVs and Association of Terrestrial Base Stations for Fronthaul Connectivity}

% ====================================================================
\maketitle

% === ABSTRACT ====================================================================
% =================================================================================
\begin{abstract}
The mushroom growth of cellular users requires novel advancements in the existing cellular infrastructure. One way to handle such a tremendous increase is to densely deploy terrestrial small-cell base stations (TSBSs) with careful management of smart backhaul/fronthaul networks. Nevertheless, terrestrial backhaul hubs significantly suffer from the dense fading environment and are difficult to install in a typical urban environment. Therefore, this paper considers the idea of replacing terrestrial backhaul network with an aerial network consisting of unmanned aerial vehicles (UAVs) to provide the fronthaul connectivity between the TSBSs and the ground core-network (GCN). To this end, we focus on the joint positioning of UAVs and the association of TSBSs such that the sum-rate of the overall system is maximized. In particular, the association problem of TSBSs with UAVs is formulated under communication-related constraints, i.e., bandwidth, number of connections to a UAV, power limit, interference threshold, UAV heights, and backhaul data rate. To meet this joint objective, we take advantage of the genetic algorithm (GA) due to the offline nature of our optimization problem. The performance of the proposed approach is evaluated using the unsupervised learning-based k-means clustering algorithm. We observe that the proposed approach is highly effective to satisfy the requirements of smart fronthaul networks.

\end{abstract}

% === KEYWORDS ====================================================================
% =================================================================================
\begin{IEEEkeywords}
Backhaul capacity, bandwidth allocation, drones, evolutionary computing, fronthaul/backhaul network, genetic algorithm (GA), optimal deployment, small cell base stations, unsupervised learning, unmanned aerial vehicles (UAVs), 5G.
\end{IEEEkeywords}

% \fi
\IEEEpeerreviewmaketitle

% ====================================================================
% ====================================================================
% ====================================================================

% === I. INTRODUCTION =============================================================
% =================================================================================
\section{Introduction}
\IEEEPARstart{T}{he} unabated growth of global cellular users and the technological advancements have drawn the attention of researchers towards novel wireless communication techniques. Therefore, fifth-generation (5G) and beyond (B5G) communication technology would be looking to introduce various kinds of network facilities in different parts of communications systems \cite{new_R3}. For the sake of simplicity, in this article, we use the generic term 5G+ to represent 5G and B5G communication standards. 5G+ communication technologies require maximum coverage and bandwidth by keeping in view the power and cost constraint of various network entities. As far as the maximum bandwidth requirement is concerned, various wireless technologies, for instance, free-space optics (FSO) and millimeter-Wave (mmWave) technology, have already been introduced \cite{fasterfiber, new_R4}. The positive side of these technologies is that they can provide hundreds of megahertz (MHz) bandwidth with the downside of limited communication range. On the other side, the staggering growth of cellular users and their immense data rate demands (e.g., video calling) leads to the concept of using terrestrial small-cell base stations (TSBSs) and has been considered as the alpha and omega of 5G+ communication technology \cite{online2, new_R5}. These TSBSs can be deployed in a range of every 250\,meters or so to make a small-cell (e.g., pico or femto), with the benefit of providing a smaller coverage area; thus, improving the data rates and reliability of cellular users.
\par In the ultra-dense deployment of TSBSs, fronthaul (the connection between baseband and radio unit) link requires a high throughput (in tens of gigabit per second (Gbps)) with sub-milliseconds latency \cite{online3}. This huge requirement of throughput with low latency can only be tackled by using fiber communication technology with a drawback of high capital expenditure (CAPEX) cost and deployment time \cite{online4}, \cite{cost}. To overcome this problem, high-performance wireless links are the best candidates to replace the fiber-based communication links. Therefore, FSO and mmWave technologies are the leading candidates as they come up with the exact requirements, which are needed in 5G+ networks. The major drawbacks of these technologies are: the requirement of line-of-sight (LoS) communication paths, sensitivity to weather conditions, and short communication range. To mitigate these drawbacks, microwave technology \cite{online3}, is the best candidate in non-LoS (NLoS) environment and is capable of covering a wide area as compared to FSO/mmWave technology; however, the downside is low data rate. To subdue aforementioned technologies, \cite{nfpidea} provided a scalable idea of using drones, unmanned aerial vehicles (UAVs), unmanned balloons, or helikite as fronthaul hubs between the TSBSs and ground core-network (GCN). To vanquish the limitations of few available NLoS fronthaul links for terrestrial users, UAVs provide the availability of wireless LoS fronthaul links, which are capable of utilizing FSO, mmWave, or radio frequency (RF) technology. Thus, terrestrial backhaul network can be replaced by an aerial network with the aim of saving CAPEX cost and deployment time.
\par UAVs technology emerged as a new paradigm shift towards 5G+ networks because of their instant, flexible, autonomous, cost-effective deployment, adaptive altitude and relocation characteristics \cite{uav1, uav2, new_R9}. In addition, UAVs can provide better LoS communication than terrestrial base stations by adjusting their altitudes intelligently. Therefore, because of their inherent characteristics, UAVs are being used in a variety of applications. For example, Google’s Loon project and Facebook’s Internet-delivery drones are practical examples to provide Internet connectivity to poorly covered areas or locations where there is no internet connectivity \cite{zuckerberg}. Considering these practical examples, UAVs can be used as aerial-hubs to provide the connectivity between TSBSs and the GCN \cite{nfpidea}. These UAVs are classified into three categories based on the altitude levels such as low altitude platforms (LAPs), medium altitude platforms (MAPs) and high altitude platforms (HAPs). They hover at an altitude from a few hundred meters to a few kilometers (up to a maximum of 20\,kms) depending on the coverage area, users location, and weather conditions. Nevertheless, there are several challenges associated with this kind of network such as TSBSs association (serving a group of network entities, e.g., TSBSs) with the UAVs, air-to-ground (ATG) channel modeling, deployment, and hover time optimization of UAVs \cite{karamUL, uavch1, uavch2, uavch3, new_R1}. This work focuses on the joint positioning of UAVs and the association of TSBSs by considering multiple communication-related constraints.
\subsection{Related Work}
In \cite{atg}, an ATG path-loss model is presented for the communication between UAVs (of LAPs category) and the ground users. Afterwards, a closed-form expression of the model was presented in \cite{optimalaltitude}. Further, considering the fixed path-loss, a closed-form expression was analytically derived for a single UAV to find the optimal altitude to maximize the coverage. Moreover, considering the parameters such as height and distance between the two UAVs, the coverage area is optimized in \cite{clouddrone}. The problem of UAVs placement and/or association of users is addressed by a few researchers in the literature \cite{10, 11, 12, 13, 14}.  All of these papers consider the UAVs as aerial base stations to provide connectivity to ground cellular users. The authors of \cite{10} addressed the placement of a single UAV as a base station by considering signal-to-noise-ratio (SNR), which is a heterogeneous quality-of-service (QoS) parameter. In \cite{11}, a comprehensive analysis for the association of ground users with a single UAV is done by considering backhaul data rate, bandwidth limit, and maximum path-loss as communication constraints. The downside of the work in \cite{10} and \cite{11}, is their impractical approach of using exhaustive search in the domain of UAV-communication for the online optimization problem, in which it is assumed to have no or incomplete knowledge of future events \cite{online_optimization}. \cite{12} addressed the association of terrestrial users with multiple UAVs, and then particle swarm optimization (PSO) algorithm is used to place the UAVs; nonetheless, PSO-based UAV placement demands a higher number of UAVs to meet users' satisfaction \cite{jointGAassociation}. The authors of \cite{13} and \cite{14} use the idea of optimal packing and game theory for the deployment of multiple UAVs. All the same, \cite{13} and\cite{14} only consider SINR criteria to solve the optimization problem.
\par In \cite{jointGAassociation}, joint UAV positioning and association of cellular users are addressed using evolutionary computing; however, the downside is again of using a computationally complex approach for the online optimization problem. Also, only bandwidth allocation constraint is considered, backhaul constraint, links availability at UAVs, etc., are not considered, which are of pivotal importance for the association and placement of UAVs. Further, energy efficiency and height constraint are ignored for UAV positioning. Similarly, \cite{19} addressed the three-dimensional ($3$D) placement of a single UAV as an aerial base station by dividing the problem into two phases, i.e., first finding the optimal altitude and then making use of the circle placement problem to find the two-dimensional ($2$D) optimized location with the objective of covering maximum users in a region. Furthermore, a commonly known evolutionary algorithm, i.e., genetic algorithm (GA) \cite{evolutionary}, \cite{arsalan}, is used to optimize UAVs trajectory by the authors of \cite{gatrajectory}. Nevertheless, the work in \cite{gatrajectory} does not take into account the communication between the UAVs and cellular users. Finally, \cite{uav3} considers the association of ground cellular users with UAVs by taking the hover time of a UAV into account. Nonetheless, the placement of UAV to improve the requirement of cellular users is not addressed.
\begin{figure*}
\begin{center}

  \includegraphics[width=0.95\textwidth%,height=7cm
  ]{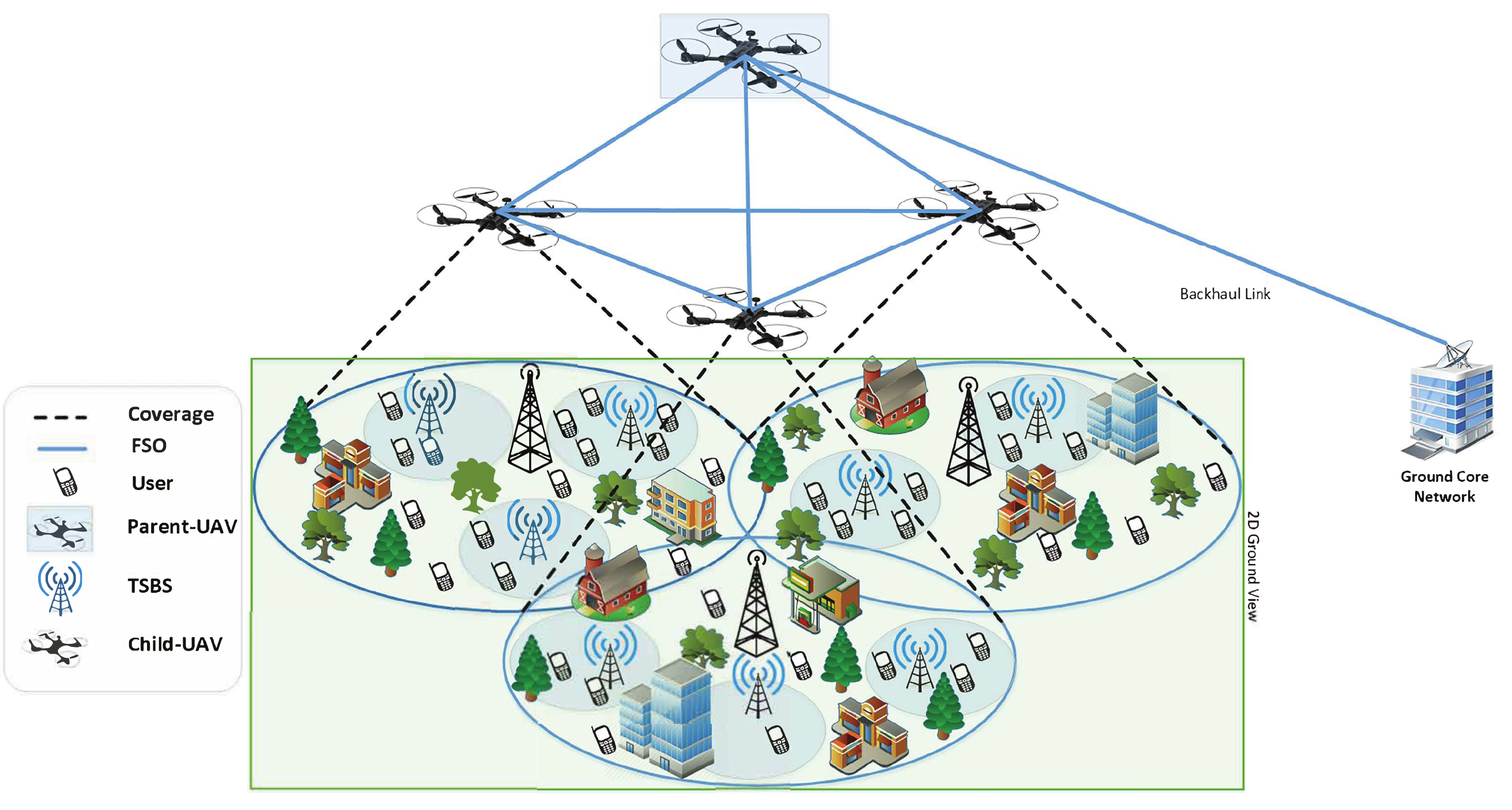}
  \caption{\textcolor{black}{Pictorial representation of UAV-enabled fronthaul network with backhaul consideration.}}\label{systemmodel}

\end{center}
\end{figure*}
\par Within the context of our work, i.e., use of UAVs as aerial-hubs to provide the fronthaul connectivity between the TSBSs and GCN, \cite{awais1} addressed the placement of UAVs using \textit{Matern type-I} hard-core process. Then the association of TSBSs is done by considering the fixed position of UAVs. A similar approach has also been used by the same authors to solve the association problem \cite{awais2}, \cite{awais3}. However, this random deployment of UAVs can cause huge interference problem and coverage issues. Also, the available resources are not efficiently utilized, and the height of all UAVs is assumed fixed. Motivated by these issues, \cite{karamUL} used the idea of deploying UAVs using the unsupervised learning-based $k\text{-means}$ clustering algorithm and then the association of TSBSs is performed, which resulted in consuming less bandwidth, achieving high sum-rate and link utilization. Nonetheless,  $k\text{-means}$ clusters the TSBSs based on the Euclidean distance and then associates to UAVs; thus, $k\text{-means}$ ignores the channel conditions, bandwidth/power allocation, objective function’s maximization or minimization, which are of supreme importance for the deployment of UAVs. Besides, the work presented in \cite{awais1, awais2, awais3, karamUL} do not consider energy efficiency and height adjustment of UAVs. To deal with this, \cite{new_R1} addressed one-dimensional ($1$D), i.e., height, placement of UAVs in the backhaul networks. All the same, in \cite{awais1, awais2, awais3, karamUL, new_R1}, the association of TSBSs with UAVs is not evaluated by searching (in $3$D) the multiple locations of UAVs, instead a fixed location of UAVs is assumed, and then the heuristic approach is used to solve the association problem. \subsection{Contribution}
\textcolor{black}{The main contributions of this work are given as follows.
\begin{enumerate}
    \item To the best of our knowledge, within the context of employing UAVs as hubs between TSBSs and GCN, this is the first work, which considers the association problem along with the $3$D positioning of UAVs. Therefore, this paper focuses on the joint positioning of UAVs and the association of TSBSs such that the sum-rate of the overall network can be maximized.
    \item The spectral efficiency of the network is considered as pivotal importance, and therefore, an algorithm is developed to serve the TSBS, which demands the highest spectral efficiency.
    \item The objective, i.e., joint placement of UAVs and association of TSBSs, is imposed by outlining stringent communication constraints, i.e., maximum available bandwidth that a UAV can distribute among its candidates, maximum number of links a UAV can support to satisfy the constraint of maximum number of carried transceivers, interference consideration between UAVs and TSBSs, maximum power at which a UAV can transmit a signal to improve energy efficiency, optimal transmit power of a UAV, optimal altitude adjustment of each UAV, and maximum backhaul data rate limit to satisfy the quality of backhaul link.
    \item The proposed optimization problem is solved using GA to achieve the objectives. The intelligence of biologically inspired selection process enables GA to outperform benchmark scheme and meet the requirements of 5G+ cellular networks. Simulation results corroborate that GA-based solutions are able to achieve the desired objective.
 %   \item A practical stochastic geometry approach is considered for the distribution of TSBSs such that a minimum distance (i.e., $250$ meters) can be maintained between them to meet the requirements for practical deployment of TSBSs \cite{online2}.
   % \item A modified algorithm to consider backhaul link constraint is presented to enhance fairness in the overall network.
\end{enumerate}
}
%Keeping in view these constraints, the resulted optimization problem is multi-constrained and complex, thus, providing room for exploring metaheuristic techniques, which are able to find near optimal solutions in short span of time. Therefore, in this work, we use GA to solve the joint problem of positioning of UAVs and the association of TSBSs with the objective function of maximizing the sum-rate of the overall system. The intelligence of biologically inspired selection process enables GA to outperform benchmark scheme and meet the requirements of 5G+ cellular networks. Simulation results confirm that GA-based solutions are able to achieve the desired objective.
% === II. System Model ========================
% =============================================
The rest of the paper is organized as follows: In Section\,\ref{system_model}, system model is presented. In Section\,\ref{problem}, problem is formulated by considering multiple communication constraints, and proposed approach is addressed in Section\,\ref{proposed_approach}. Practical deployment aspects of the proposed approach are given in Section\,\ref{deployment}. Simulation-based results are presented in Section\,\ref{results}. Finally, in Section\,\ref{conclusion}, conclusion is drawn with potential future directions.
\section{System Model}\label{system_model}

% =======
% FIG. 01
% =======

Fig.\,\ref{systemmodel} depicts a dense urban environment of heterogeneous network $(\text{HetNet})$, where five network entities, i.e., UAVs, GCN, mobile cellular users, and TSBSs overlaid on a macro base station, are shown. The pictorial representation delineates a 5G+ network, which aims to provide wireless backhaul/fronthaul connectivity to TSBSs via UAV-hubs (for brevity, we refer these UAV-hubs as child-UAVs in the rest of the article). In the considered diagram, TSBSs (e.g., pico and femto) are deployed in a geographical area, where they are aggregating and routing the downlink\footnote{Also, capable of routing uplink traffic. In this paper, we only assumed downlink traffic.} traffic of cellular users via child-UAVs to the GCN. In addition, child-UAVs are hovering at altitude $h_{j}$, where $j$ represents the $j^{th}$ child-UAV, to provide wireless fronthaul connectivity between TSBSs and GCN. These child-UAVs are hovering\footnote{in a range of few 100 meters to a few kilometers, the typical range is around 20\,km, which is due to the reason of higher probability of LoS at higher altitude. However, in our work, we have assumed a range of 300\,m to 800\,m for the sake of minimum path-loss.} autonomously, which are controlled by another UAV, named as parent-UAV\footnote{\textcolor{black}{The purpose of using parent-UAV is to avoid multiple backhaul links. In addition, to control the swarm of child-UAVs, parent-UAV has been considered.}}. \textcolor{black}{Further, parent-UAV is acting as a communication-hub between the child-UAVs and the GCN.} Furthermore, parent-UAV is flying at an altitude higher than child-UAVs to provide a perfect $\text{LoS}$ communication to child-UAVs and the GCN. Also, the communication link from child-UAVs to parent-UAV and from parent-UAV to GCN is based on FSO technology\footnote{Here, it is considered that there are no losses in this link. Therefore, perfect $\text{LoS}$ is maintained. Nevertheless, communication losses can be taken into account, and we leave this as future direction \cite{nfpidea}.}, which aims to provide high-speed wireless connection \cite{new_R8}. Additionally, child-UAVs have the capability of exchanging the control information (e.g., signal-to-interference-plus-noise ratio $(\text{SINR})$ between each child-UAV and TSBS, data rate, and bandwidth requirements of TSBSs) with each other and the parent-UAV. Nevertheless, every child-UAV is responsible for sharing the accumulated information directly to parent-UAV, where parent-UAV takes care of backhaul link's limitations. \textcolor{black}{Moreover, the connection between the child-UAVs and TSBSs is established using an RF communication link, which is sub-6GHz.}
\par For the sake of notational convenience, we assume the location of TSBSs and child-UAVs as $\mathbf{q}_{i}\in\mathbb{O}^{2}=(v_{i}, w_{i})$ and $\mathbf{z}_{j}\in\mathbb{O}^{3}=(x_{j},y_{j}, h_{j})$, respectively, where $ i=\{1, 2, 3, ..., T\}$, $j =\{1, 2, 3, ..., U\}$, and $\mathbb{O}^{d}$ represents the Euclidean space in $d$ dimensions. Without loss of generality, let $\mathbf{Q}$ and $\mathbf{Z}$ be the set of location of all TSBSs and the child-UAVs, respectively, such that $ \mathbf{Q}=\{\mathbf{q}_1, \mathbf{q}_2, \mathbf{q}_3, ..., \mathbf{q}_T\}$ and $ \mathbf{Z}=\{\mathbf{z}_1, \mathbf{z}_2, \mathbf{z}_3, ..., \mathbf{z}_U\}$. \textcolor{black}{To describe the path-loss model and the association problem of TSBSs with child-UAVs, the initial position of child-UAVs is kept as random, i.e., by considering the range of deployment area and height limits of child-UAVs, the child-UAVs are randomly deployed in the predefined area. This is due to the reason that having a totally random initial position of child-UAVs provide an opportunity for a more dispersed search
in the entire solution space; thus, avoiding the problem of premature convergence in the GA, which we deal with later in this paper.} However, it is important to note that the distribution of child-UAVs is random just to illustrate the problem, and we will address the positioning of child-UAVs later in this paper. On the other side, the distribution of TSBSs is obtained using a \textit{Matern type-I} hard-core process \cite{Matern} with the average density of $\delta/m^2$; therefore, the resulting process gives the average number of TSBSs as
\begin{equation}\label{matern}
T^{\text{avg}}= \delta\cdot\exp(-\delta\pi D_{\text{min}}^2)\cdot\bigtriangleup\: \:,
\end{equation}
where $\bigtriangleup$ denotes the area in which the TSBSs are deployed. Additionally, $D_{\text{min}}$ represents the minimum distance (separation) between the two TSBSs. It can be said that average number of TSBSs in a region is the product of average density, $\delta$, and the total area, $\bigtriangleup$. Further, we have summarized the distribution of TSBSs and child-UAVs in Algorithm\,\ref{algo1}. For the sake of understanding, the input parameters of Algorithm\,\ref{algo1} are described in Table\,\ref{table1}.
\begin{algorithm}[t!]
 \label{algo1}
\DontPrintSemicolon
  \KwInput{$\bigtriangleup$, $\delta$, $D_{\text{min}}$, $U$, $h_{j}$}
  \KwOutput{{$(v_{i}, w_{i})$} and $(x_{j}, y_{j}, h_{j})$}
  \textbf{Distribution of TSBSs:}\;
  Apply \textit{Matern type-I} hard-core process by taking $\bigtriangleup$, $\delta$, and $D_{min}$ as input parameters\;
  \textit{Matern} $(\bigtriangleup$,\,$\delta$,\,$D_{min})\:\Rightarrow\:$ $(v_{i},w_{i})$\;
   \textbf{Initial Distribution of child-UAVs:}\;
  Apply a MATLAB function, named \textit{Random}, by taking $\bigtriangleup$, $U$, and $h_{j}$ as input parameters\;
   {\tcp{\textcolor{black}{$\textit{Random}$ MATLAB function considers the upper and lower bounds (e.g., height, area, etc.) of child-UAVs placement}} }
  $\textit{Random}$ $(\bigtriangleup$, $U$, $h_j)\:\Rightarrow\:$ $(x_{j},y_{j}, h_{j})$
\caption{\textcolor{black}{Distribution of TSBSs and initial child-UAVs placement}}
\end{algorithm}
\begin{figure}
\centering
\includegraphics[scale=0.35]{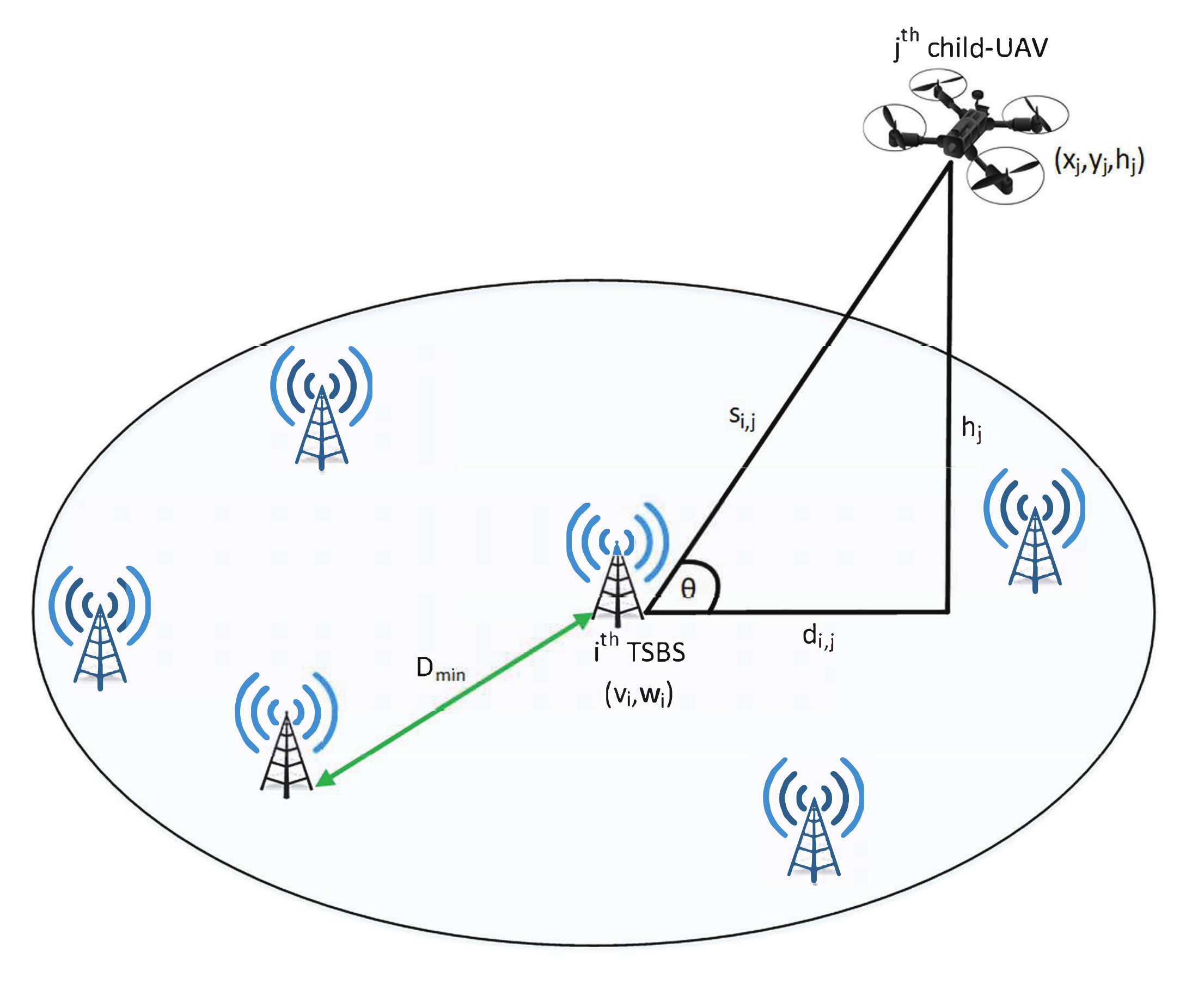}
\caption{\textcolor{black}{Graphical view of the system model for communication of a child-UAV with a TSBS.}}
\label{fig1stochastic}
\end{figure}
\par Considering the distribution of TSBSs and child-UAVs using Algorithm\,\ref{algo1}, the horizontal distance between the $i^{th}$ TSBS and the $j^{th}$ child-UAV is calculated as
\begin{equation}\label{eq:1}
{d}_{{i},{j}}= \sqrt{{(v_i-x_{j})}^2+{(w_i-y_{j})}^2}\: \:.
\end{equation}
\par The angle of elevation (in degrees) from $i^{th}$ TSBS to $j^{th}$ child-UAV is denoted as
\begin{equation}\label{eq:2}
\theta_{i, j}=\arctan\Big(\frac{h_{j}}{d_{i,j}}\Big)\: \:,
\end{equation}
where $h_{j}$ is the height of $j^{th}$ child-UAV. Also, considering the stochastic geometry shown in Fig.\,\ref{fig1stochastic}, the vertical distance between the $i^{th}$ TSBS and the $j^{th}$ child-UAV can be written as
\begin{equation}\label{eq:3}
{s}_{{i},{j}}= \sqrt{{h}^2_{j}+d^{2}_{i,j}}\: \:.
\end{equation}
\subsection{Path-loss Model}
To calculate path-loss between the child-UAVs and TSBSs, a widely used air-to-ground ($\text{ATG}$) path-loss model is used, which is presented in \cite{atg} and \cite{optimalaltitude}. The authors considered two categories of communication links, i.e., $\text{LoS}$ and $\text{NLoS}$, where $\text{NLoS}$ path takes reflection and diffraction phenomena into account. Also, probability of $\text{LoS}$ communication link, which plays a vital role in UAV communication is formulated in \cite{atg} and \cite{optimalaltitude} by considering Equations \eqref{eq:1}\,$\text{-}$\,\eqref{eq:3} as
\begin{equation}\label{eq:4}
{\varrho}^{L}_{i,j} = \frac{1}{1+\alpha\cdot\exp\big\{-\beta(\theta_{i, j}-\alpha)\big\}}\:\:,
\end{equation}
where ${\varrho}^{L}_{i,j}$ represents the probability of $\text{LoS}$ between the $i^{th}$ TSBS and the $j^{th}$ child-UAV, $\alpha$ and $\beta$ are the environment-dependent (e.g., urban, suburban, and rural) constants. In a nutshell, the above equation portrays that probability of $\text{LoS}$, ${\varrho}^{L}$, is highly dependent on height and environment. Additionally, considering Equation \eqref{eq:4}, probability of $\text{NLoS}$ between $i^{th}$ TSBS and $j^{th}$ child-UAV is given by
\begin{equation}\label{eq:5}
{\varrho}^{N}_{i,j}=1-{\varrho}^{L}_{i,j}\:\:.
\end{equation}
\par Finally, the $\text{ATG}$ path-loss model derived using the above equations is presented as
\begin{equation}\label{eq:6}
\Gamma_{{i},{j}} (\text{dB}) = {F^{0}_{i,j} + {\varrho}^{L}_{i,j}\cdot\xi^{L}+{\varrho}^{N}_{i,j}\cdot\xi^{N}} \: \:,
\end{equation}
where $\xi^{L}$ and $\xi^{N}$ depict the efficiency (or attenuation factors) of $\text{LoS}$ and $\text{NLoS}$ communication paths, respectively. Furthermore, $F^{0}$ is the free-space path-loss $(\text{FSPL})$ and is calculated between $i^{th}$ TSBS and $j^{th}$ child-UAV as
\begin{equation}\label{eq:7}
F^{0}_{i,j} (\text{dB}) =\gamma\cdot\log_{10}\Big(\frac{4\pi\cdot s_{{i},{j}}}{\lambda_{\text{carrier}}}\Big)\: \: ,
\end{equation}
where $\gamma$ is the path-loss exponent and $\lambda_{\text{carrier}}$ can be written as
\begin{equation}\label{eq:8}
\lambda_{\text{carrier}}=\Big(\frac{c}{f_{\text{carrier}}}\Big)\: \:,
\end{equation}
where $c$ and $f_{\text{carrier}}$ are the speed of light and carrier frequency, respectively.
\subsection{Optimal Transmit Power and $\text{SINR}$ Calculation}
The optimal transmit power is calculated by applying an interference constraint. Assume $a_{i,j}\in\{0, 1\}$ represents the connectivity between $i^{th}$ TSBS and $j^{th}$ child-UAV. If a connection establishes between $i^{th}$ TSBS and $j^{th}$ child-UAV, then value is 1, i.e.,  $a_{i,j}=1$, and zero otherwise. Also, assume $I_{th}$ denotes the interference threshold for the $i^{th}$ TSBS, therefore, mathematically
\begin{equation}\label{eq:9}
    {a_{i, j}}{ \cdot g_{i, j} \cdot{\Omega}_{i, j}} \leq I_{th} \:, \qquad\forall{i, j},
\end{equation}
where $g_{i, j}$ is product of the magnitude squared of the channel gain and the inverse of the path-loss between the $i^{th}$ TSBS and the $j^{th}$ child-UAV. In addition, ${\Omega}_{i, j}$ denotes the optimal transmit power of the $j^{th}$ child-UAV towards $i^{th}$ TSBS with the constraint given by
\begin{equation}\label{eq:10}
   {\Psi}^{\text{min}}\leq{\Omega}_{i, j} \leq {\Psi}^{\text{max}} \:, \qquad \forall{i, j} \:,
\end{equation}
where ${\Psi}^{{\text{min}}}$ and ${\Psi}^{{\text{max}}}$ are the minimum and maximum transmit power of a child-UAV, respectively.
\par Therefore, by using path-loss given in \eqref{eq:6} and the optimal transmit power, $\Omega$, received power, ${p}^{r}$, at $i^{th}$ TSBS from $j^{th}$ child-UAV is calculated as
\begin{equation}\label{eq:11}
{p}^{r}_{i,j} (\text{dB})=  10\cdot\log_{10}({\Omega}_{i, j})+\wp_{i, j}-\Gamma_{{i},{j}}\:\:,
\end{equation}
where we have also considered the fading, $\wp$, to show typical urban environment, and is formulated as
\begin{equation}\label{eq:12}
\wp_{i,j} (\text{dB})={\varrho}^{L}_{i,j}\cdot\zeta_{0}+{\varrho}^{N}_{i,j}\cdot\zeta_{1} \:,
\end{equation}
where the envelopes of $\zeta_{0}$ and $\zeta_{1}$ are \textit{Nakagami}
distributed, $|\zeta_{\iota}| \sim Nakagami(m)$, here $\iota=\{0, 1\}$, and $m$ is the shape parameter, which takes the value 1 and 4 for Rayleigh and Rician fading, respectively.
\par Thus, by taking Equation \eqref{eq:11}, $\text{SINR}$, $\Im$, at $i^{th}$ TSBS is obtained as
\begin{equation}\label{eq:13}
\Im_{i,j} =\frac{{p}^{r'}_{i,j}}{\sigma_{n}^{2}+I_{s}}\:\:,
\end{equation}
where ${p}^{r'}_{i,j}$ is the received power (in Watts) at $i^{th}$ TSBS from $j^{th}$ child-UAV and $\sigma_n^{2}$ is the noise power of that particular link. Further, due to the assumption of omnidirectional antennas used at child-UAVs, $I_{s}$ denotes the sum of interference received from other $(U-1)$ child-UAVs.
\section{Problem Formulation}\label{problem}
The reliable and efficient communication of cellular users with TSBSs and GCN is considerably dependent on the fronthaul link of child-UAVs. Therefore, optimal placement of child-UAVs and association of TSBSs is of pivotal importance as it can be viable for throughput maximization, better connectivity, and $\text{QoS}$ experience.      \par Consider the downlink scenario of the network, where TSBSs are downloading the data from GCN via child-UAVs. In this context, the optimal positioning of child-UAVs and the association of TSBSs with these child-UAVs is a challenging task. Specifically, stringent communication-related constraints put a strong bound towards the joint objective, i.e.,  the association of TSBSs with child-UAVs and the optimal location of child-UAVs to maximize the sum-rate of the overall network. Therefore, below we first describe the communication constraints between the child-UAVs and TSBS and between the parent-UAV and the GCN. Later on, we formulate the objective of our proposed work.
\subsection{Communication Constraints}
\textcolor{black}{Consider the distribution of TSBSs and the fixed location of child-UAVs, which is summarized in Algorithm\,\ref{algo1}. For a fixed location of child-UAVs, the communication between the TSBSs and child-UAVs is limited by a number of factors, which are discussed below.}
\begin{enumerate}
            \item A child-UAV cannot transmit beyond a maximum power, ${\Psi}^{{\text{max}}}$, to improve energy efficiency (defined later in this section).
            \item An interference constraint, $I_{th}$, is also considered to find the optimal transmit power of a child-UAV.
            \item To avoid overloading, a child-UAV, $j$, can accommodate a maximum of $\ell^{\text{max}}$ links.
          \item There is a maximum bandwidth limit, $B$, which a child-UAV can distribute among its candidates (TSBSs).
          \item To maintain
          $\text{QoS}$ requirements, minimum SINR, $\Im^{\text{min}}$, criteria is taken into account as it is of pivotal importance for the distribution of bandwidth.
          \item Finally, an $i^{th}$ TSBS will only be served by a particular $j^{th}$ child-UAV.
\end{enumerate}
Besides, to maintain the quality of the backhhaul link, which is the communication path between the parent-UAV and the GCN, a maximum backhaul data rate limit, $R_{B}$, is considered.
\subsection{Objective Formulation}
Keeping in view the aforementioned communication constraints, the objective of this work is to find the optimal positioning of child-UAVs and the best possible association  of TSBSs to child-UAVs such that the sum-rate of the overall system can be maximized. The association of TSBSs to child-UAVs is dependent on communication factors, i.e., ${\Psi}^{{\text{max}}}$, $I_{th}$, $\ell^{\text{max}}$, $B$, $\Im^{\text{min}}$, and $R_{B}$. Let $\mathbf{A}$, with dimension $T\times U$, be the association matrix, where the rows and columns of matrix $\mathbf{A}$ represent the TSBSs and child-UAVs, respectively. Also, $a_{i,j}\in\{0, 1\}$ be the $(i, j)^{th}$ entry of matrix $\mathbf{A}$. Hence, the joint objective, i.e., optimal positioning of child-UAVs $\mathbf{Z}^{*}$ and the association of TSBSs to child-UAVs $\mathbf{A}^{*}$ is, therefore, mathematically formulated as
\begin{align}
 \begin{split}\label{fitness_function}
 \mathbf{Z}^{*},\mathbf{A}^{*}=\operatorname*{argmax}_{\mathbf{Z}\in \mathbb{O}^{3\times U}, a_{i,j}\in\{0,1\}} {\sum_{i=1}^{{T}}}\sum_{j=1}^{{U}} {r}_{i,j}^{d}\cdot a_{i,j}
\end{split}\\
\shortintertext{s.t.}
%\begin{split}
\label{band}
 & {\sum_{i=1}^{{T}}} {{b}_{i,j}}.a_{i,j} \leq {B}_{j},  &\forall&{j}
%\end{split}
\\
%\begin{split}
\label{links}
&{\sum_{i=1}^{{T}}} a_{i,j} \leq \ell_j^{\text{max}},  &\forall&{j}
%\end{split}
\\
%\begin{split}
\label{power}
&\Psi_j^{\text{min}}\leq\Omega_{{i,j}} \leq {\Psi}_{j}^{\text{max}}, &\forall&{j}
%\end{split}\\
\\
%\begin{split}
\label{interference}
&     {a_{i, j}}{ \cdot g_{i, j} \cdot{\Omega}_{i, j}} \leq I_{th}, &\forall&{i, j}&
%\end{split}
\\
%\begin{split}
\label{heightconstraint}
& {h_{\text{min}}\leq h_j\leq h_{\text{max}}}, &\forall&{j}&
 %\end{split}
 \\
%\begin{split}
\label{SINR}
&\Im_{i,j} \cdot a_{i,j} \geq  \Im^{\text{min}}, &\forall&{i, j}&
%\end{split}
\\
%\begin{split}
\label{connection}
&{\sum_{j=1}^{{U}}} a_{i,j} \leq 1, &\forall&{i}&
%\end{split}
\\
%\begin{split}
\label{bac_rate}
&{\sum_{i=1}^{{T}}}\sum_{j=1}^{{U}} {r}_{i,j}^{d}\cdot a_{i,j} \leq {R}_{B},&&
%\end{split}
\end{align}
\textcolor{black}{where $a_{i,j}$ and $\mathbf{Z}$ are the optimization parameters, which represent the connectivity of a TSBS with child-UAVs and the position of child-UAVs, respectively. Further, ${r}_{i,j}^{d}$ is the demanded data rate of $i^{th}$ TSBS from $j^{th}$ child-UAV. From Equation \eqref{fitness_function},
\begin{equation}\label{fitness}
F_{s}={\sum_{i=1}^{{T}}}\sum_{j=1}^{{U}} {r}_{i,j}^{d}\cdot a_{i,j}\:,
\end{equation}
where $F_{s}$ represents the total achieved sum-rate from the overall network and we call this as fitness function for optimal search of child-UAVs via $\text{GA}$ (explained in next section). }
\par \textcolor{black}{In constraint \eqref{band}, $B_{j}$ is the maximum available bandwidth to $j^{th}$ child-UAV, which it can distribute among the candidates, i.e., TSBSs, and $b_{i,j}$ is the required bandwidth of $i^{th}$ TSBS from $j^{th}$ child-UAV, which is calculated as
\begin{equation}\label{eqbw}
{b}_{i,j}=\frac{r_{i,j}^{d}}{\log_{2}{(1+{\Im}_{i,j})}}\:.
\end{equation}}
\par \textcolor{black}{Additionally, constraint \eqref{links} and \eqref{power} denote the maximum number of links a child-UAV can support (or, in other words, maximum number of TSBSs, which a child-UAV can serve), and maximum \& minimum power at which a child-UAV can transmit the initialization signal, respectively. Also, constraint \eqref{interference} considers interference threshold, constraint \eqref{heightconstraint} limits the height of child-UAV in a specific range $(h_{\text{min}}, h_{\text{max}})$, and constraint \eqref{SINR} takes care of minimum $\text{SINR}$ criterion to maintain $\text{QoS}$ requirements as it can play a pivotal role in the distribution of bandwidth. Furthermore, constraint \eqref{connection} restricts a TSBS to be connected with one particular child-UAV; thus, one TSBS will only be served by one child-UAV. Moreover, backhaul data rate limit, $R_{B}$, which is the communication link between the parent-UAV and GCN, is represented in constraint \eqref{bac_rate}.}
\par Finally, having the sum-rate value from Equation \eqref{fitness}, we define the energy efficiency $(E_{\text{eff}})$ of the overall network, in $\text{bits/s/Watt}$, as
\begin{equation}\label{eqee}
  E_{\text{eff}}=\frac{{\sum\limits_{i=1}^{T}}\sum\limits_{j=1}^{U} {r}_{i,j}^{d}\cdot{a_{i,j}}}{P_{\text{Total}}}\: \:,
\end{equation}
where $P_{\text{Total}}$ is the total transmit power and is written as
\begin{equation}
    P_{\text{Total}}=\epsilon\cdot{\sum\limits_{i=1}^{T}\sum\limits_{j=1}^{U} {\Omega}_{i, j}\cdot a_{i, j}
 + K \times \rho_{i,j}^{c} }\:\:,
\end{equation}
where $K$ is the total number of associated TSBSs, where $K\leq{T}$. Also, $\rho_{i,j}^{c}$ is the circuit power, cost by the link between an $i^{th}$ TSBS and $j^{th}$ child-UAV. Moreover, $\epsilon$ denotes the inverse of power amplifier efficiency.
\section{Proposed Approach Using GA}\label{proposed_approach}

\subsection{Association of TSBSs with child-UAVs}
The sole purpose of addressing an association algorithm is to find out which TSBS will be served by which particular child-UAV. In addition, based on the demanded data rate, ${r}^{d}$, of a TSBS, how the bandwidth allocation can be done such that child-UAV can efficiently utilize available bandwidth resource. Additionally, considering the limited number of carrier transceivers to not overload a child-UAV, how the available number of links at a child-UAV can intelligently be employed. Further, how $\text{QoS}$ can be maintained by employing an interference constraint, furthermore, how intelligent picking of TSBSs can be done such that the backhaul link can support the demanded capacity. Furthermore, how energy efficiency, $E_{\text{eff}}$, can be maximized by finding out the optimal transmit power, $\Omega$, of a child-UAV. Finally, how fairness can be introduced such that most of the TSBSs in the network shall remain connected. Keeping in view these questions, an association algorithm is addressed in the following section with details. This algorithm is addressed by keeping in view the distribution of child-UAVs, and TSBSs mentioned in Algorithm\,\ref{algo1}; however, the association algorithm is integrated with the proposed child-UAVs placement discussed in Section \ref{proposed GA}.
\par Following two steps for the association of TSBSs are to be carried out with reference to child-UAVs. These two steps are summarized in Algorithm\,\ref{algo2}.
\begin{algorithm}[t!]
\label{algo2}
\DontPrintSemicolon
  \KwInput{$T$, $U$, $\ell^{\text{max}}$, $B$,  $b_{i,j}$, $r_{i,j} $, $\Im_{i,j}$, ${R}_{B}$}
  \KwOutput{{$ {\mathbf{A}}$}}
  {\tcp{\textcolor{black}{Initialize matrix $ {\mathbf{A}}$ with all zeros}} }
  \KwInitialize{$ {\mathbf{A}}=\varnothing$}
    {\tcp{\textcolor{black}{Step at each TSBS}} }
  \For{$i=1$ \textbf{to} ${T}$}
  {
 \textcolor{black}{Select child-UAV from which an $i^{th}$ TSBS receives maximum $\text{SINR}$, $\Im^\text{max}$. Also, $i^{th}$ TSBS validates if it satisfies the constraint \eqref{SINR}, then TSBS updates  $a_{i,j}=1$. Otherwise,
 $i^{th}$ TSBS updates $a_{i,j}=0$.\;}
  }
    {\tcp{\textcolor{black}{Step at each child-UAV}} }
    \For{$j=1$ \textbf{to} ${U}$}
  {
  \textbf{Initialize counters:} $ {C_{\ell}}=0$, $ {C_{b}}=0$
  \newline
     \While{ $ {C_{\ell}} < {\ell^{\text{max}}} \wedge  {C_{b}} <  B$}
     {
     \text{Find highest spectral efficient TSBS\;}
   {

             \If{$ {C_{b}} + {b_{i,j}} \leq  B$}
    {

        Update $ {C_{\ell}}$ =$  {C_{\ell}} + 1$ and $ {C_{b}}$ = $  {C_{b}}$ + ${b_{i,j}}$\;
                        \Else
    {
Update ${a_{i, j}} = 0$,\;

    }

        }

   }
   }
  }
\caption{\textcolor{black}{Association of TSBSs with child-UAVs}}
\end{algorithm}
\begin{itemize}
\item Assuming the positions of child-UAVs and TSBSs discussed in Algorithm\,\ref{algo1}, each child-UAV broadcasts an initialization signal with a transmit power $\Omega$ (by considering the constraint \eqref{power} and \eqref{interference}), which is received by all the TSBSs deployed on ground (due to the assumption of omnidirectional antennas). \textcolor{black}{Therefore, each TSBS calculates the $\text{SINR}$, $\Im$, by using Equation \eqref{eq:13}. Next, a TSBS verifies the constraint \eqref{SINR}, i.e., $\Im^{\text{max}}>\Im^{\text{min}}$. Later on, if the constraint \eqref{SINR} is satisfied, then a TSBS sends the feedback $1$ to that particular child-UAV from which it receives the maximum SINR, $\Im^{\text{max}}$, and null vector to the remaining child-UAVs (where each entry corresponds to non-selected child-UAV). Otherwise, it sends feedback $0$ to all the child-UAVs.} Mathematically, it can be assumed that in matrix $\mathbf{A}$, we have a number of non-zero entries in each column and only one non-zero entry (if $\Im^{\text{max}}>\Im^{\text{min}}$) in each row. This shows that if $i^{th}$ TSBS is connected with $j^{th}$ child-UAV, then there is value $1$ in that particular row $(i^{th})$ and column $(j^{th})$. Further, rest of the entries in $i^{th}$ row are zero, which shows that a single TSBS is only connected with one child-UAV; thus, satisfy the constraint \eqref{connection}. This step is summarized in lines $1$-$3$ of Algorithm\,\ref{algo2}.
\item In the second step, each child-UAV goes through its list, and it starts allocating the bandwidth using Equation \eqref{eqbw} until the available bandwidth resource ends at each child-UAV, therefore, each child-UAV keeps check on constraint \eqref{band}. It is important to note that child-UAV firstly allocates bandwidth to that particular TSBS, which is highly spectral efficient, $\gimel_{\text{SE}}$, and is written as
\begin{equation}\label{se}
	\gimel_{\text{SE}}=\frac{r_{i,j}^{d}}{b_{i,j}}\:.
\end{equation}
Next, each child-UAV also makes sure to not overload itself by keeping track of a maximum number of links that it can accommodate; thus, each child-UAV keeps constraint \eqref{links} in view. If any of the constraint, i.e., maximum allowed bandwidth, $B$, or the maximum number of links, $\ell^{\text{max}}$, does not satisfy, then a child-UAV sends feedback zero to that particular TSBS showing the indicator that it cannot serve; thereby, child-UAV moves to the next TSBS to serve (depending on the availability of resources, i.e., $\ell^{\text{max}}$ and $B$). Further, if all the requests of that particular child-UAV are entertained, then the task of that child-UAV is considered as completed. Therefore, to this point, each child-UAV has done the job of getting maximum sum-rate by giving priority to highly spectral efficient TSBSs. The pseudo-code of this step can be seen in lines $3$-$9$ of Algorithm\,\ref{algo2}.
\par Mathematically, we can say that we have an updated matrix $\mathbf{A}$, which may or may not have the same number of ones (as in the previous step (lines $1$-$3$)), and it depends on the communication constraint \eqref{band} and \eqref{links}. The step $2$ (lines $3$-$9$) is to be executed at each child-UAV independently; thus, the distributive approach can be used to save the run time of the algorithm. Thus far sum-rate of the algorithm has been maximized by keeping in lieu all the constraints except constraint \eqref{bac_rate}, which is explained below and is summarized in Algorithm\,\ref{algo3}.
\begin{algorithm}[t!]
\label{algo3}
\DontPrintSemicolon
 {\tcp{\textcolor{black}{Task at parent-UAV}} }
   \KwInitialize{$ {F_{s}}$ as total sum-rate of associated TSBSs}
    \While{$ {F_{s}}$ $>$ ${R}_{B}$}
   {
   Select child-UAV with max. associated TSBSs $\triangleright$  {\tcp{\textcolor{black}{This approach will introduce fairness for all regions due to the reason of selection (may or may not) of different child-UAV in each iteration}}}
Select TSBS with minimum data rate, $min$ $(r_{i,j}^d)$, demand\;
De-associate the selected pair ($i^{th}$ TSBS of $j^{th}$ child-UAV) and update $a_{i, j} = 0$,
$ {C_{\ell}}$ = $ {C_{\ell}} - 1$, $ {F_{s}}$ = $ {F_{s}} -$ $r_{i,j}^d$ and $ {C_{b}}$ = $ {C_{b}} -$ $b_{i,j}$ \;
   }
\caption{Algorithm for constraint \eqref{bac_rate}}
\end{algorithm}
\item In the final step, all the child-UAVs propel the aggregated information (i.e., SINR, demanded rate, etc.) to parent-UAV, where parent-UAV takes care of the backhaul link capacity. To ensure backhaul link capacity, parent-UAV sums, $(F_{s})$, all the demanded data rates (of the associated TSBSs with child-UAVs) and verifies if the achieved sum-rate $(F_{s})$ is within the backhaul link’s capacity $(R_{B})$ or not. If $F_{s}$ is within the backhaul rate limit, then the association algorithm completes. Otherwise, to satisfy backhaul link capacity, parent-UAV selects the child-UAV which has associated maximum number of TSBSs with itself, and within the list of selected child-UAV, parent-UAV picks that particular TSBS, which is demanding minimum data rate, $min$ $(r_{i,j}^d)$. After picking the particular pair (child-UAV and its respective TSBS), parent-UAV de-associates the request of that particular TSBS by updating the respective entry of matrix $\mathbf{A}$ to zero, i.e., $a_{i,j}=0$. And, after each de-association, parent-UAV checks constraint \eqref{bac_rate}, if the constraint is still not satisfied, then parent-UAV selects the entry of the same child-UAV and searches for TSBS with minimum demanded rate and de-associates its request. If achieved sum-rate, $F_{s}$, is still not within the backhaul data rate limit, then parent-UAV selects the next child-UAV (or the same one if it has still maximum associated TSBSs) with maximum associated TSBSs, and repeats the same process until the constraint \eqref{bac_rate} is not satisfied. This final step, to be executed at parent-UAV, is summarized in Algorithm\,\ref{algo3}.
\end{itemize}
\par The sole purpose of selecting the child-UAV with maximum associated requests is to introduce fairness to all the TSBSs within the network. For example, suppose the minimum associated child-UAV is selected. In that case, it might serve one or two TSBSs, which would be a wrong approach in terms of energy efficiency, UAV usage, and fairness to other TSBSs in the network etc. Besides, the purpose of selecting TSBS for de-association with minimum data rate demand is due to the aim of maximizing the sum-rate of the overall network.
\subsection{Child-UAV Positioning} \label{proposed GA}
Inspired by natural genes and the selection process, GA evolved as a new technique for numerical optimization of many problems. In simple words, GA has been introduced as a fast search optimization algorithm, which is beneficial in most computationally complex optimization problems. Therefore, considering the offline-nature (due to the fixed position of TSBSs) of our optimization problem, we use the idea of GA \cite{evolutionary} for the optimum placement of child-UAVs and association of TSBSs, with the objective of maximizing the fitness function (sum-rate of the overall system) defined in \eqref{fitness}.
Different from the traditional GA, we propose a modified implementation of GA, i.e., without the use of crossover operation. Such implementation is adopted to tailor GA according to the requirements of the child-UAV positioning problem. Within the context of child-UAV positioning, every child-UAV tends to provide coverage to its surrounding area. Hence, sharing the coordinates of a child-UAV with others (to generate new offspring) leads GA to produce many unfeasible solutions, increasing the time complexity. Therefore, for the reproduction of new generation, we propose \emph{mutation-only GA} to solve the problem. Below we describe a detailed description of this approach.
\subsubsection{GA's Initial Population}
GA consists of an initial population represented by the set ${\mathcal{N}}$ and $\mathbf{Z}_{e}$ is the element of the set and it represents a row vector, which contains the $3$D location of all child-UAVs, i.e., $\mathbf{Z}_{e} = [\mathbf{z}_{1}, \mathbf{z}_{2}, \mathbf{z}_{3}, ..., \mathbf{z}_{U}]$. Also, $e\in\{1, 2, 3, ..., S \}$, where $S$ denotes the total size of population. Thus, $\mathcal{N} \triangleq [\mathbf{Z}_{1}, \mathbf{Z}_{2}, \mathbf{Z}_{3}, ..., \mathbf{Z}_{S}]^\mathsf{T}$, where $[\cdot]^\mathsf{T}$ depicts the transpose operator.
\par In terms of $\text{GA}$, $\mathbf{Z}_{e}$ represents an individual chromosome having length $3\times{U}$(\text{total number of genes ($g_{T}$}), where $g_{T}=\{g_{1}, g_{2}, g_{3}, ..., g_{3\times{U}}\})$, where numeric value $3$ shows $(x,y,h)$ location of a child-UAV and $U$ is the total number of child-UAVs. Therefore, in a nutshell, each chromosome, $\mathbf{Z}_{e}$, contains the $(x,y,h)$ location of all child-UAVs, and there are a total of $S$ chromosomes in the entire population.
\par As part of this study, we came up with trying the different flavors of GA's population, i.e., population generated by $k\text{-means}$ clustering algorithm, a totally random population (given in Algorithm\,\ref{algo1}), and hybrid of both. Various experiments showed that random population (for child-UAVs positioning) generated using Algorithm\,\ref{algo1} is clearly the winner as it provides an opportunity for a more dispersed search in the entire solution space; thus, avoiding the problem of premature convergence. However, addressing the results of different flavors of GA's population is not the focus of our paper. Therefore, we describe and analyze GA using a random population in the rest of this article.
\par The initial phase of GA is to generate a random population of size $S\times{g_{T}}$ and then calculate the fitness value (sum-rate ($F_{s}$)) against each chromosome, $\mathbf{Z}_{e}$, using the association algorithms mentioned in Algorithm\,\ref{algo2} and \ref{algo3} (as our objective is to jointly optimize the positions of child-UAVs and association of TSBSs with these child-UAVs). Till this point, we have a fitness value corresponding to an individual chromosome, and our objective is to maximize this fitness value by keeping in view Algorithm\,\ref{algo2} and \ref{algo3}. Therefore, below we present the updated generation (G+) based on the fitness value of an individual chromosome.
\subsubsection{GA's Updated Population}\label{updated_pop}
In the new generation (G+), we select $40\%$ best chromosomes (based on their fitness value) from the previous generation, and we call these chromosomes elite parents. Additionally, the selection of the remaining 60\% chromosomes is based on roulette wheel selection (RWS), which selects the new $60\%$ chromosomes (from the previous generation) depending on their fitness value such that a chromosome with a higher fitness value has a higher probability of selection, e.g., elite parents have the higher probability of selection even though dreg parents (chromosomes having low fitness value) can also be selected with a low selection probability.
\par
\begin{figure}
\centering
\includegraphics[scale=0.52]{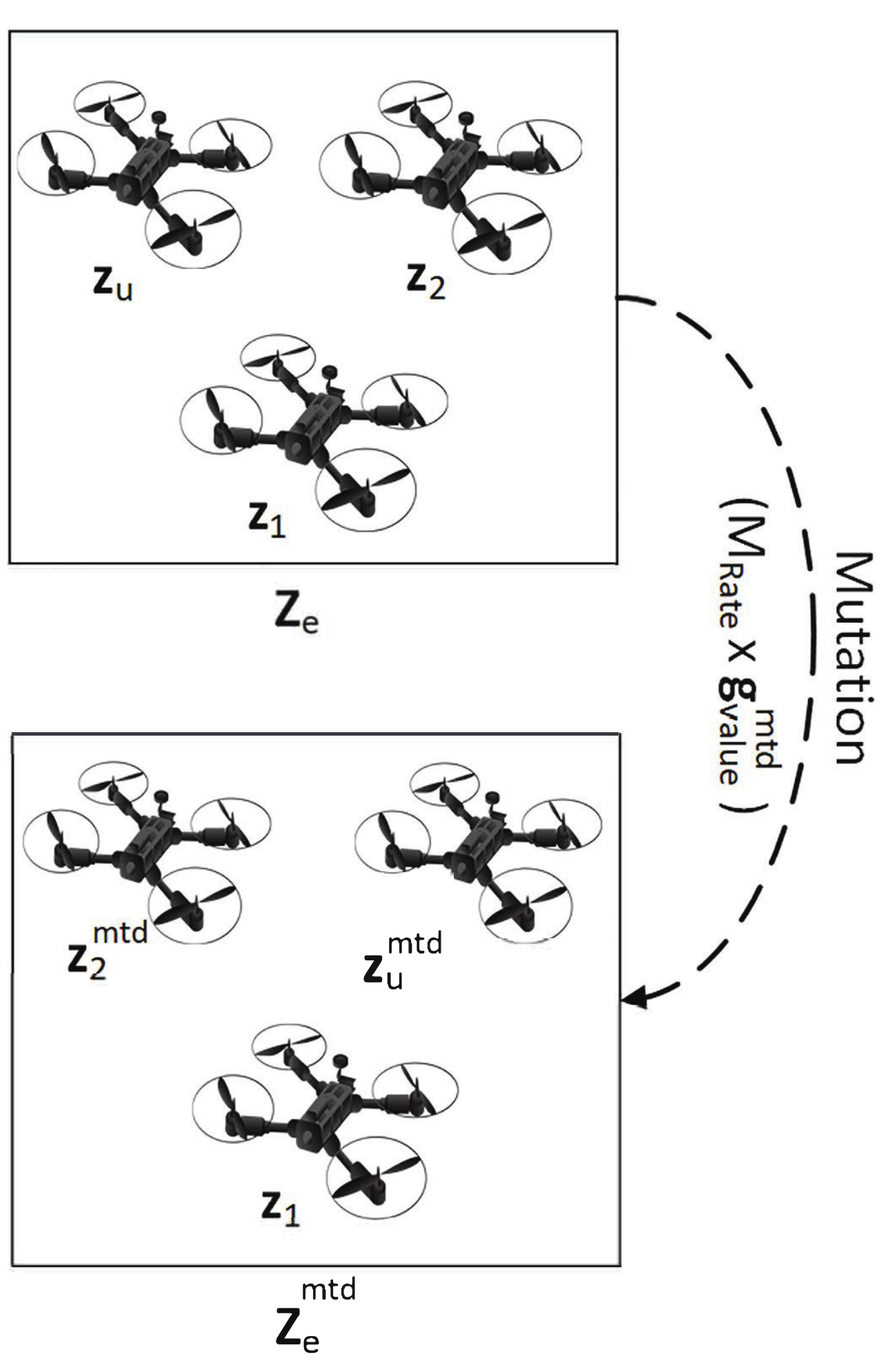}
\caption{\textcolor{black}{Pictorial representation of updated child-UAVs positions via proposed GA: mutation operation.}}
\label{figmutation}
\end{figure}

 After the selection of new $60\%$ chromosomes, mutation operation inherent to all genetic algorithms, is applied to these newly selected chromosomes. The mutation operation is the process in which the position of a single or multiple child-UAVs (depending on number of genes being mutated of a chromosome) is updated. In our model, mutation operation is applied in the following manner:
\begin{equation}\label{mutation}
    \mathbf{Z}_e^{\text{mtd}}=\mathbf{Z}_{e} \pm (M_{\text{Rate}}\times{\mathbf{g}_{\text{value}}^{\text{mtd}}})\:\:,
\end{equation}
where $M_{\text{Rate}}$ is the mutation rate and $\mathbf{g}_{\text{value}}^{\text{mtd}}$ is a vector of gene values (from a preselected chromosome) selected for mutation. It is important to note that the selection of a gene value is totally random. Also, the addition or subtraction of term $(M_{\text{Rate}}\times{\mathbf{g}_{\text{value}}^{\text{mtd}}})$ in a preselected chromosome is dependent on a coin toss, if its head then the value would be added, and subtracted otherwise. Pictorially, the principle of mutation is drawn in Fig.\,\ref{figmutation}, where one or multiple child-UAVs (depending on vector size being mutated) relocate their positions in the predefined area of operation.
\par In summary, the updated population (G+), is the combination of $40\%$ original chromosomes (elite parents) and $60\%$ mutated chromosomes (mutated parents).
\subsubsection{Calculation of $F_{s}$ at G+}
Finally, having the updated population, the fitness function, i.e., maximizing the sum-rate of the overall network, is reevaluated against individual chromosome $(\mathbf{Z}_{e})$ using Algorithm\,\ref{algo2} and \ref{algo3},
and if no improvement is observed after some generations, then GA stops, otherwise it selects a new generation using Section \ref{updated_pop} and reevaluates the fitness function.
\par The pseudo-code of GA is summarized in Algorithm\,\ref{algo4}. The algorithm takes $h_{\text{min}},\, h_{\text{max}},\, \bigtriangleup,\, S,$ and $U$ as input parameters and it gives the fittest chromosome $\mathbf{Z}^{*}$ and matrix $\mathbf{A}^{*}$, i.e., optimal position of child-UAVs and best possible association of TSBSs, respectively, as output.
\begin{algorithm} \label{algo4}
\DontPrintSemicolon
  \KwInput{$h_{\text{min}}$,\,$h_{\text{max}}$,\,$\bigtriangleup$,\,$S$,\,$U$}
  \KwOutput{{$ {\mathbf{Z}^{*}}$\, ${\mathbf{A}^{*}}$}}
Create an initial population using a MATLAB function: $InitialPop (h_{\text{min}}, h_{\text{max}}, \bigtriangleup, S, U )$  \;
Associate TSBSs (using Algorithm\,\ref{algo2} and \ref{algo3}) against each chromosome of initial population\;
Calculate $F_{s}$ using \eqref{fitness} for each chromosome

  \For{$t=2$ \textbf{to} ${G_{it}}$}
  {
 Select $40\%$ elite parents (based on max. $F_{s}$ value) from previous population\;
 Select $60\%$ new chromosomes from previous population using the approach discussed in Section \ref{updated_pop}\;
 Apply mutation using Equation \eqref{mutation} on $60\%$ newly selected chromosomes\;
 Apply Algorithm\,\ref{algo2} and \ref{algo3} on updated population ($40\%$ elite parents and $60\%$ mutated parents) for the association of TSBSs\;
 Calculate $F_{s}$ using \eqref{fitness} against each chromosome\;
 \If{$ no\:change\:in\:F_s\:in\:5\:generations$}
    {
       {\textbf{break for loop}}\;
         Select the chromosome $\mathbf{Z}^{*}$ and association matrix $\mathbf{A}^{*}$ that gives best $F_{s}$
        }

 }

\caption{\textcolor{black}{Pseudo-Code of GA for Child-UAVs Positioning and Association of TSBSs}}
\end{algorithm}
\section{Practical Deployment of GA} \label{deployment}
To implement our proposed approach, the location of TSBSs and their demanded data rates $(r^{d})$, which will be provided by mobile network operators $(\text{MNOs})$, are required. Further, the operational parameters, for instance, $\alpha$, $\beta$, etc., are also needed to calculate the path-loss and to assign bandwidth to TSBSs. Having these details, below we address the practical implementation aspects of the proposed approach.
\par Considering the scenario of our system, there are two possible ways of running GA for the joint positioning of child-UAVs and the association of TSBSs. One of the possible ways is to implement GA at particular network entities, i.e., GCN, baseband unit (BBU), assuming the availability of Cloud-Radio Access Network (CRAN), or at TSBSs due to their static position. This approach would not cause any problem in terms of power consumption as the aforementioned static entities contain enough operating power and viable for GA's running. Moreover, in our work, positioning of child-UAVs and association of TSBSs is needed to be determined once in a time due to the static nature of our ground network entities unless or until a new TSBS has been deployed in the network by $\text{MNOs}$, therefore, GA is a promising approach in such environment.
\par An alternative of implementing GA is at movable entities, i.e., child-UAVs and parent-UAV, due to their on-demand, flexible, autonomous deployment and low latency characteristics towards the association of TSBSs. Furthermore, as the child-UAVs will only be sharing the control information (i.e., $\text{SINR}$) with each other, therefore, GA's implementation can be viable. However, running GA at child-UAVs may reduce the hover time of a child-UAV \cite{new_R2, uav3}; thereby, a child-UAV cannot stay aloft for a longer time period \cite{new_R1}. Secondly, in our network, child-UAVs will regularly be replaced with new ones; hence, running GA at child-UAVs may not be a feasible approach from the perspective of battery life. Moreover, the same limitations are of parent-UAV.
In a nutshell, practical implementation of GA is favorable at static network entities of our system model and is of pivotal importance towards power saving of the overall system.
\section{Simulation Environment and Analysis of Results}\label{results}
In this section, we firstly describe the simulation environment. Later on, a detailed analysis of obtained results is presented.
\begin{table} {}
\caption{Simulation parameters \cite{optimalaltitude, karamUL}.}
\centering
 \begin{tabular}{c c c}
 \hline
 \textbf{Parameter} & \textbf{Value} & \textbf{Description}  \\
 \hline
 %\hline
 $f_{\text{carrier}}$ & 2\,GHz & Carrier frequency\\ [1ex]
% \hline
 $\alpha$,  $\beta$ & 9.61, 0.16 & Environment constants \\
 %\hline
 $\xi^{L}$, $\xi^{N}$& \{1, 20\}\,dB & Efficiency of $\text{LoS}$ \& $\text{NLoS}$ links\\
 %\hline

 %\hline
 $\delta$ & $2\times10^{-6}/\,\text{m}^2$  & Density of TSBSs\\
 %\hline
  $\gamma$ & 2 & Path-loss exponent\\
 %\hline
 $\bigtriangleup$ & 16\,$\text{km}^{2}$&Total area \\
 $\Psi^{\text{max}}$ & 1.3\,Watt&Max. transmission power\\
 $\sigma_n^{2}$ & -125\,dB&Noise power \\
 $\ell^{\text{max}}$ & 7&Max. no. of links \\
 $B$ & 200\,MHz&Max. bandwidth limit \\
 $\Im^{\text{min}}$ & -10\,dB&Min. SINR level \\
 $R_{B}$ & 1.66\,Gbps&Backhaul data rate limit\\
 $\rho^{c}$ & 0.1\,Watt&Circuit power \\
     $S$ & 50&Population size \\
 $M_{\text{Rate}}$ & 9\%&Mutation rate \\
  ${\rm{g}^{\text{mtd}}}$ & 4&Mutated genes \\
    ${G_{it}}$ & 50&Max. iterations of GA \\
 \hline
\end{tabular}
\label{table1}
\end{table}
\begin{table*}[ht]
\caption{\textcolor{black}{Comparison of results with different evaluation parameters.}}
\centering
  \begin{tabular}{|c|c|c|}
    \hline
    {\textbf{Evaluation Parameters}} &
      {\textbf{$\mathbf{k}\text{-means}$}} &
     {\textbf{$\text{GA}$}}  \\
    \hline
    Associated TSBSs (\%)  & 76.96 & 94.53
 \\
    \hline
    Sum-Rate (Gbps)  & 1.33 & 1.61   \\
    \hline
  Average Bandwidth Consumption (MHz) &	156.28	&147.59\\
    \hline
    Energy Efficiency (Mbps/Watt) &	43.63&	43.19\\
    \hline
  \end{tabular}
  \label{table2}
\end{table*}
\subsection{Simulation Environment}
A dense urban environment has been considered, where TSBSs are distributed using a \textit{Matern type-I} hard-core process with a density of $\delta$ per meter square. Additionally, random data rate demands are assigned to deployed TSBSs from a data rate vector $\mathbf{r}^{d}=\{20, 40, 60, 80, 100\}$, where the values are in Mbps. Also, to satisfy the requirements of 5G network \cite{online2}, a minimum distance $(D_\text{min})$ of $250$\,m is maintained between the two TSBSs. In addition, four child-UAVs $(U=4)$ are considered and the minimum and maximum altitude $(h_{\text{min}}, h_{\text{max}})$ of child-UAVs is limited to $300$\,m and $800$\,m, respectively. Further, $\epsilon$ is $38$\%, and the interference threshold is limited to $1.1943\time 10^{-14}$\,Watt. Furthermore, the rest of the simulation parameters (unless stated specifically) and their description is given in Table\,\ref{table1}. Moreover, the results are averaged over $1000$ Monte Carlo realizations.
\subsection{Analysis of Results}
\begin{figure}
\centering
\includegraphics[scale=0.59]{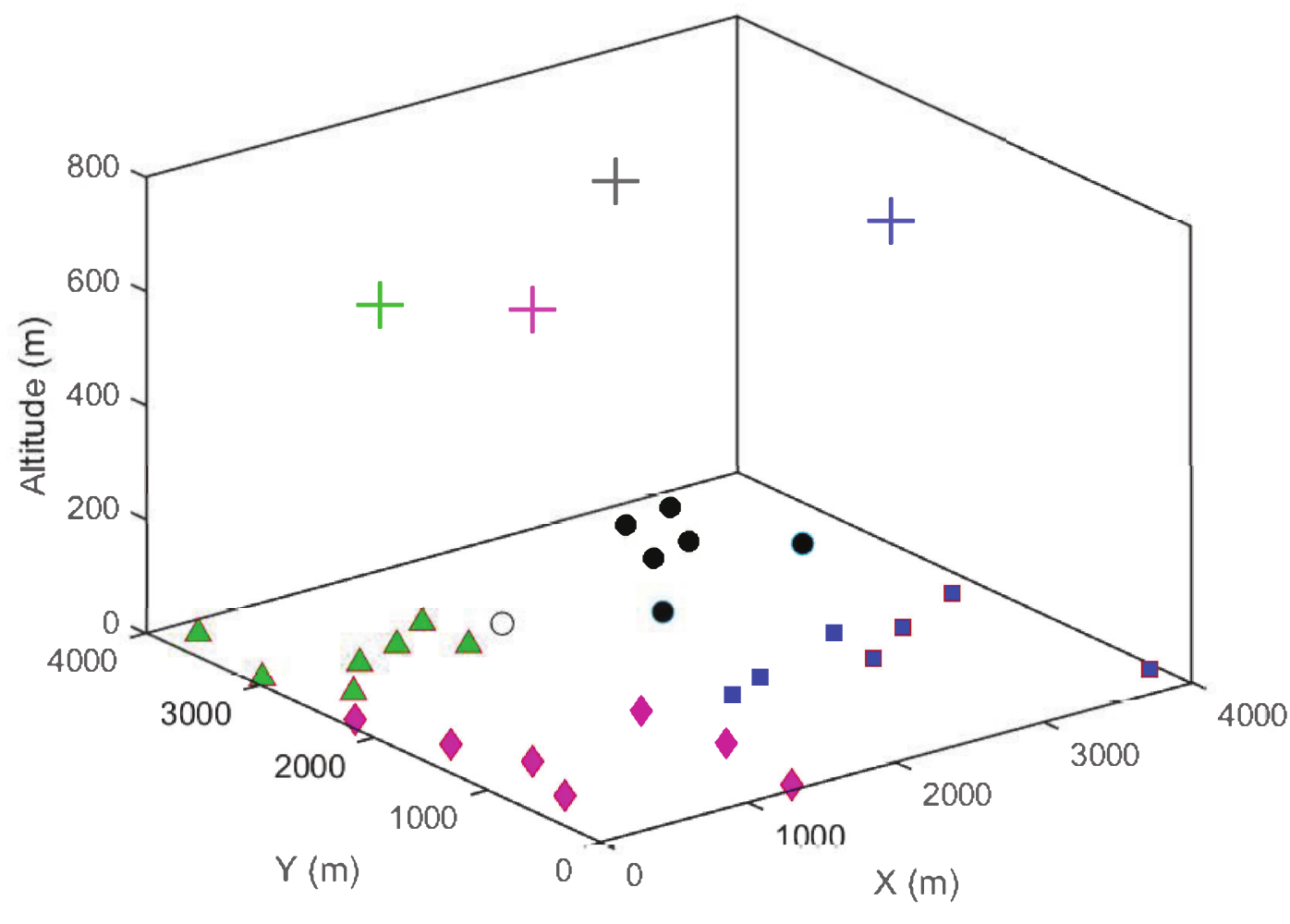}
\caption{\textcolor{black}{Optimal $3$D placement of child-UAVs using GA and the association of TSBSs. $+$ sign with different colors shows child-UAVs; filled markers are associated TSBSs with respective $+$ child-UAV; unfilled marker denotes unassociated TSBS. }}
\label{snapGA}
\end{figure}
\par Fig.\,\ref{snapGA} shows a snapshot of the association of TSBSs with child-UAVs at a particular iteration. Additionally, the optimal $3$D placement of child-UAVs is presented, which is obtained using GA. It can be observed that there is only one TSBS in an outage (or say is not served),  which is due to the reason of stringent communication constraints. In particular, in the surroundings of green child-UAV, one TSBS is not served because the constraint \eqref{links} has reached its maximum limit. Also, the unassociated TSBS is not served by other child-UAVs either due to low SINR criteria (constraint \eqref{SINR}) or high bandwidth demand (constraint \eqref{band}). Below we evaluate the performance of the proposed GA with unsupervised learning-based $k\text{-means}$ clustering algorithm.
\par
\begin{figure}
\centering
\includegraphics[scale=0.59]{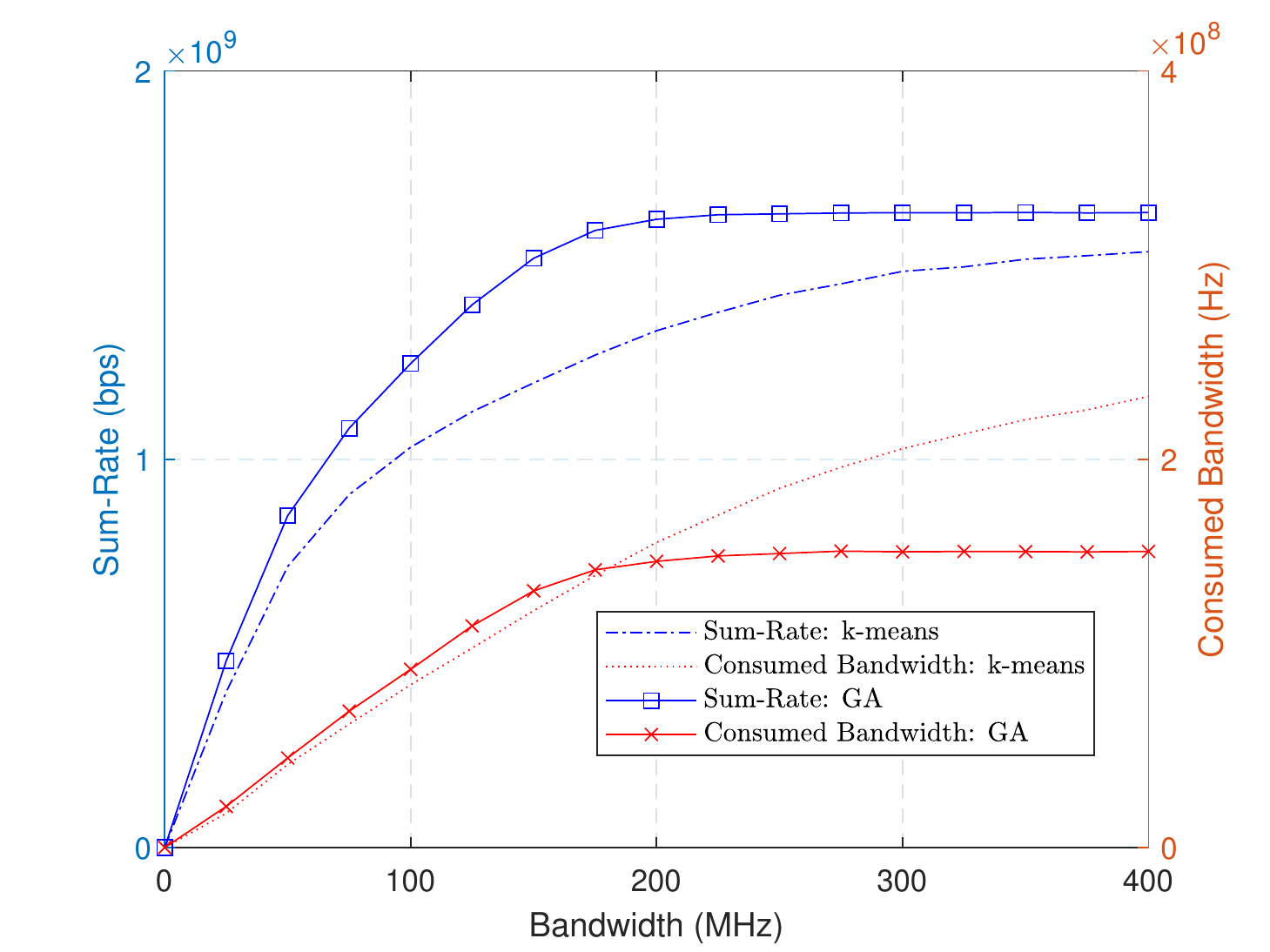}
\caption{\textcolor{black}{Analysis of sum-rate and consumed bandwidth with constraint (\ref{band}).}}
\label{yyplot}
\end{figure}

\begin{figure}
\centering
\includegraphics[scale=0.59]{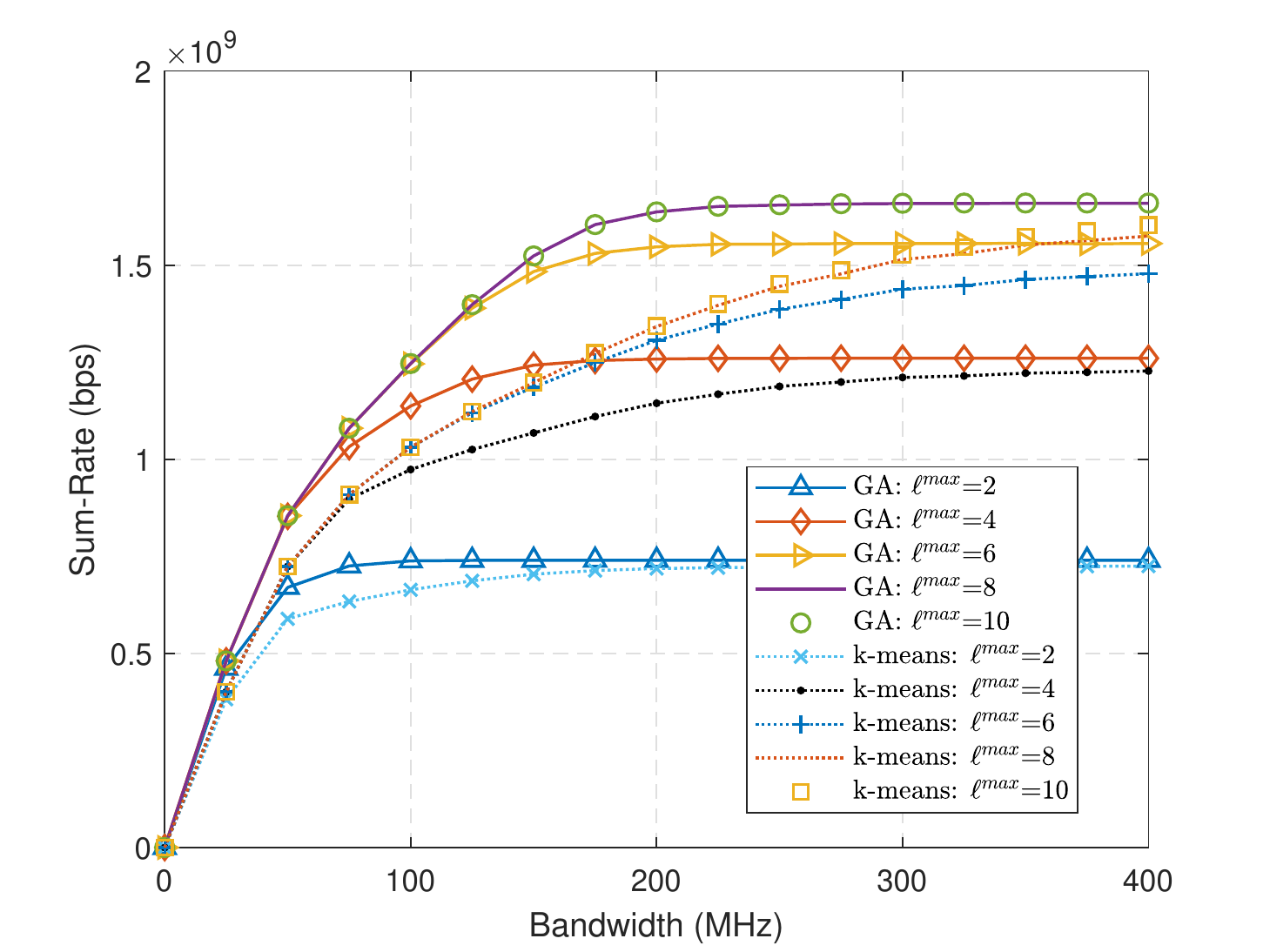}
\caption{\textcolor{black}{Comparison of sum-rate with constraint \eqref{band} and \eqref{links}.}}
\label{figlinks}
\end{figure}

\begin{figure}
\centering
\includegraphics[scale=0.59]{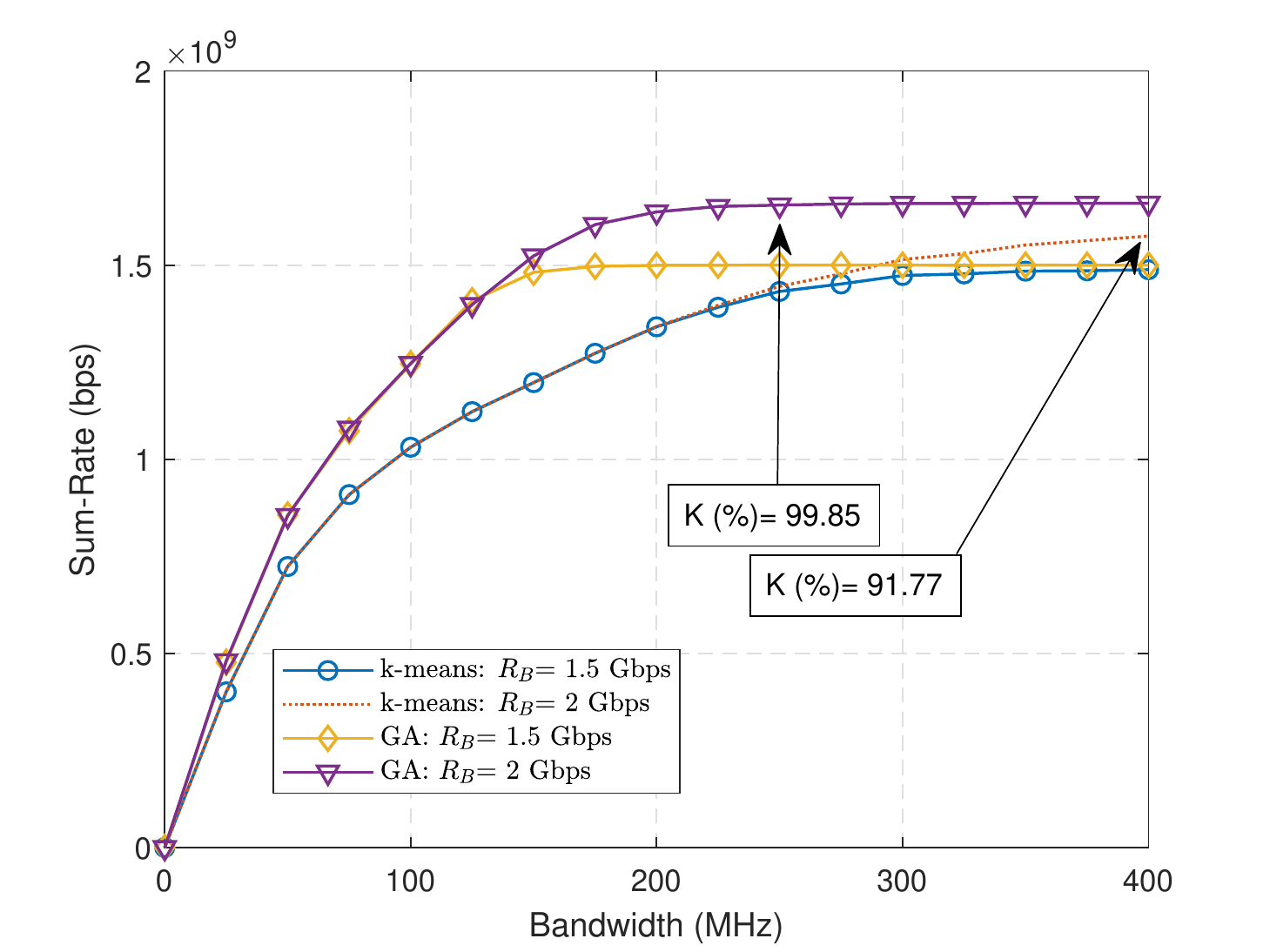}
\caption{\textcolor{black}{Comparison of sum-rate with constraint \eqref{band} and \eqref{bac_rate}.}}
\label{figback}
\end{figure}

 Table\,\ref{table2} gives a comparison of proposed approach with $k\text{-means}$. In particular, the table presents comparison for various number of evaluation parameters, i.e., percentage of associated TSBSs, sum-rate $(F_{s})$, average bandwidth consumed by child-UAVs, and achieved value of $E_{\text{eff}}$. Numerical values show that proposed GA outperforms $k\text{-means}$-based child-UAVs distribution. For example, associated TSBSs by GA are approximately $95$\% with the achieved sum-rate of $97$\% at a less bandwidth consumption of $147.59$\,MHz than $k\text{-means}$. Besides, in the case of GA, the achieved value of $E_{\text{eff}}$ is $43.19$\,Mbps/Watt, which is a little less due to serving the higher number of TSBSs; thus, the denominator of \eqref{eqee} has a higher value, which reduces the energy efficiency of GA. Nevertheless, the less value of $E_{\text{eff}}$ as compared to $k\text{-means}$ is negligible. In a nutshell, GA's performance is the best for all kind of evaluation parameters. Furthermore, it can be observed that GA's sum-rate is still a little less than the backhaul data rate limit $(R_{B})$, which we investigate in Figs.\footnote{It is important to note that the results of Figs.\,\ref{yyplot},\,\ref{figlinks}, and\,\ref{figback} are drawn by assuming the child-UAVs positions given in Fig.\,\ref{snapGA}. Thus, GA does not search for the new location of child-UAVs when the constraints are varied.}\,\ref{yyplot} and\,\ref{figlinks}. Moreover, for the sake of simplicity and the comparable performance of $k\text{-means}$ with GA, we investigate the results of GA with $k\text{-means}$, in the rest of the article.       \par \textcolor{black}{Fig.\,\ref{yyplot} reveals the performance of achieved sum-rate and the average consumed bandwidth when the available bandwidth resource is varied from $0$ to $400$\,MHz. Also, a comparison between the two schemes for child-UAVs distribution, i.e., $k\text{-means}$ and GA, is drawn. It can be observed that the sum-rate is increasing exponentially for both schemes. However, GA outperforms. Further, in the case of GA, no increase in sum-rate (approximately after $1.64$\,Gbps) can be observed after around $225$\,MHz bandwidth; thus, increasing only the constraint \eqref{band} does not improve the performance of sum-rate, and we investigate this in Fig.\,\ref{figlinks}. Nevertheless, $k\text{-means}$ keeps on increasing sum-rate when the bandwidth resource increases, which shows that $k\text{-means}$-based distribution of child-UAVs is bandwidth-hungry. Furthermore, $k\text{-means}$ performance is closed to GA at an expensive bandwidth demand of $400$\,MHz. On the other hand, the red curves depict the consumption of bandwidth for both schemes. It can be noted that GA stops consuming the bandwidth resource after $225$\,MHz, which is due to the reason that GA has achieved the desired objective. Nonetheless, on the other hand, $k\text{-means}$ keeps on consuming the bandwidth to maximize sum-rate, thus; $k\text{-means}$ scheme is bandwidth-hungry.}
\par In Fig.\,\ref{figlinks}, a comparison of GA and $k\text{-means}$ is unveiled when the bandwidth constraint is increased from $0$ to $400$\,MHz. In addition, different curves for various limits of constraint \eqref{links}, i.e., number of links supported by child-UAVs, are also portrayed. In the figure, an increasing trend of sum-rate with the increase of bandwidth resource $(B)$ and number of links $(\ell^{\text{max}})$ can be noticed for both the schemes. However, GA outperforms in both constraints, i.e., constraint \eqref{band} and \eqref{links}. For instance, in the case of GA, the combination of $\ell^{\text{max}}=8$ and $B=250$\,MHz maximizes the sum-rate, and there is no further improvement as all the TSBSs (and their required demands of data rate) are served. It is also important to note that going beyond $\ell^{\text{max}}=8$ does not make any difference for GA. Nevertheless, $k\text{-means}$ algorithm achieves the sum-rate of approximately $1.60$\,Gbps for maximum number of resources, i.e., $\ell^{\text{max}}=10$ and $B=400$\,MHz. Therefore, $k\text{-means}$ demands a higher number of constraint limits to meet $R_{B}$ or to serve all the TSBSs.

\par Fig.\,\ref{figback} depicts a comparison of achieved sum-rate when the bandwidth and backhaul data rate constraints (constraint \eqref{band} and \eqref{bac_rate}) are varied. In addition, the numerical value of percentage of associated TSBSs is written. In case of GA, it can be seen that the convergence of sum-rate starts at the available bandwidth resource of $175$\,MHz (for $R_B= 1.5$\,Gbps) and $250$\,MHz (for $R_B= 2$\,Gbps). Also, at $R_B= 2$\,Gbps, the percentage of associated TSBSs is approximately $100$\% and the sum-rate is maximized to $1.66$\,Gbps and it cannot go beyond as all the TSBSs are served and there are no further data rate demands. In contrast, sum-rate of $k\text{-means}$ converges at bandwidth of $375$\,MHz when $R_B= 1.5\,$Gbps. Also, in case of $R_B= 2\,$Gbps, $k\text{-means}$ serves around $92$\% TSBSs at maximum bandwidth constraint value, i.e., $400$\,MHz, and it demands more bandwidth resource to associate further TSBSs and to maximize sum-rate. Thus, results of Fig.\,\ref{figback} concludes that $k\text{-means}$ lags in achieving backhaul data rate limit and associates less number of TSBSs, which is due to the reason of higher bandwidth demands.
\par
\begin{figure}
\centering
\includegraphics[scale=0.59]{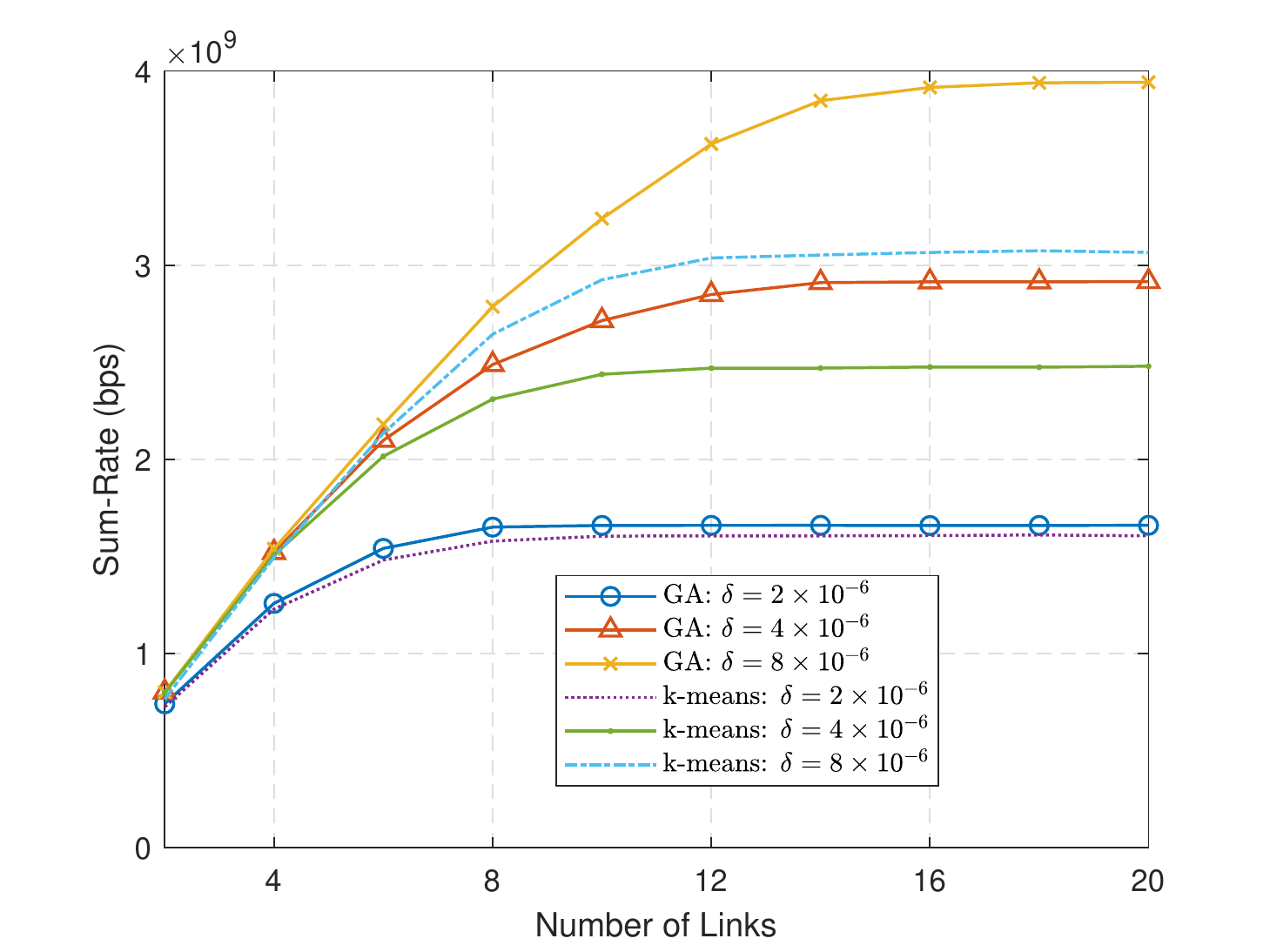}
\caption{\textcolor{black}{Evaluation of sum-rate with constraint \eqref{links} for different values of TSBSs density. $B= 400\,$MHz and $R_{B}= 4\,$Gbps.}}
\label{figsumlins}
\end{figure}

\begin{figure}
\centering
\includegraphics[scale=0.59]{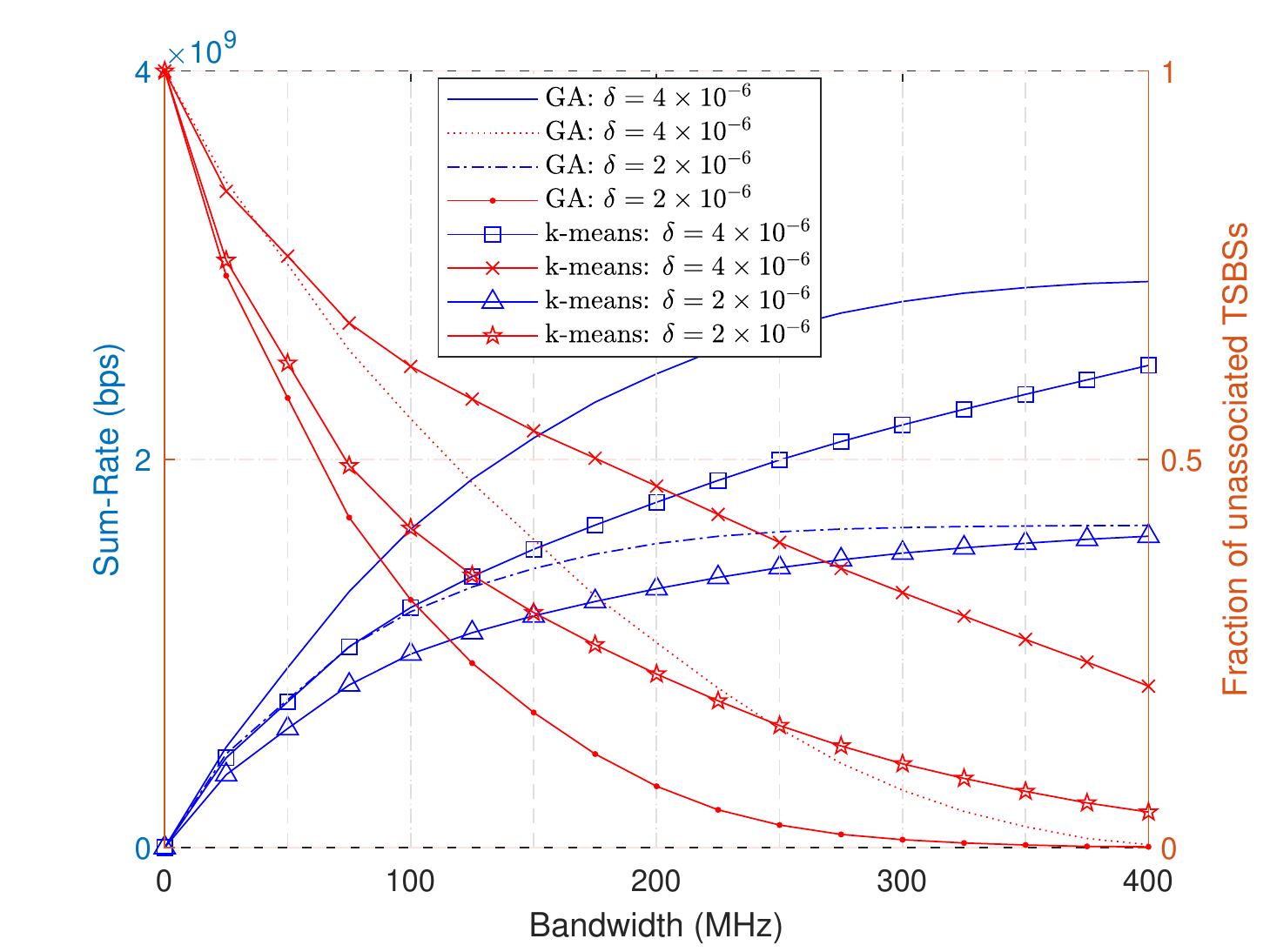}
\caption{\textcolor{black}{Analysis of sum-rate and fraction of unassociated TSBSs with constraint \eqref{band} for different values of TSBSs density. $B= 400\,$MHz, $\ell^{\text{max}}= 20$, and $R_{B}= 4\,$Gbps.}}
\label{figunassociated}
\end{figure}

 Till this point, we have analyzed the performance of proposed GA by considering the fixed network size, i.e., $\delta= 2\times10^{-6}/\,\text{m}^2$. Therefore, in Figs.\,\ref{figsumlins} and \ref{figunassociated}, comparison of results is evaluated by varying the network size. However, it is important to note that by changing the network size, optimal positions of GA are researched by using Algorithm\,\ref{algo4}, $B= 400\,$MHz, $R_{B}= 4\,$Gbps, and the parameters given in Table\,\ref{table1}. Moreover, the number of child-UAVs are kept fixed, i.e., $U= 4$. The results are drawn with different constraint values.
\par Fig.\,\ref{figsumlins} gives the performance of sum-rate when the number of links supported by child-UAVs are increased from $2$ to $20$. In addition, different curves are plotted for various values of TSBSs density $(\delta)$ and for the comparison between the two schemes, i.e., proposed GA and traditional $k\text{-means}$-based positioning of child-UAVs. Initially, for a lower value of $\delta$, performance of both the schemes is almost same; however, with the increase of $\delta$, achieved sum-rate by GA is greater and it reaches the backhaul data rate limit at $\ell^{\text{max}}= 20$ when the density of TSBSs is $8\times10^{-6}/\,\text{m}^2$. Further, in case of GA, for $2\times10^{-6}/\,\text{m}^2$ and $4\times10^{-6}/\,\text{m}^2$, sum-rate value does not increase after certain range because all the TSBSs are served at that point. On the other hand, sum-rate performance of $k\text{-means}$ is lower at $4\times10^{-6}/\,\text{m}^2$ and at $8\times10^{-6}/\,\text{m}^2$ due to the reason of serving lower number of TSBSs; thus, $k\text{-means}$ demands extra resources, for instance, bandwidth, power, etc. to increase sum-rate.
\par Finally, in Fig.\,\ref{figunassociated}, a detailed comparison (in terms of sum-rate and unassociated TSBSs) between the performance of $k\text{-means}$ and GA is drawn for different values of $\delta$. As before, $k\text{-means}$ results in achieving a low sum-rate and the number of associated TSBSs. In contrast, in the case of GA, the ratio $(K/T)$ of unassociated TSBSs is lower, and it reaches zero at $B= 400\,$MHz for both the values of $\delta$. Additionally, the achieved sum-rate by GA is maximum for both the values of $\delta$ as all the TSBSs are served; for instance, at $B= 400\,$MHz, the ratio of unassociated TSBSs is zero.
\par In a nutshell, a detailed comparison of results with various constraints and evaluation parameters shows a significant performance of the proposed GA-based distribution of child-UAVs. In particular, the achieved sum-rate and associated TSBSs are higher. The proposed GA is practically deployable considering the offline nature of our optimization function, i.e., optimal placement of child-UAVs and the association of static TSBSs.
\section{Conclusions and Future Directions}\label{conclusion}
Motivated by UAV communication, this article focused on optimal placement of UAVs as fronthaul-hubs for the backhaul connectivity of TSBSs. In particular, the joint optimal placement of UAV-hubs and the association of TSBSs is formulated so that the overall network's sum-rate can be maximized. Besides, multiple communication constraints, i.e., backhaul data rate limit, available bandwidth, power, minimum and maximum altitude, and link capacity of UAV-hubs, are considered for the association of TSBSs. To achieve such an objective with stringent communication constraints, we proposed GA for the joint optimal placement of UAV-hubs and association of TSBSs. Simulation results showed a significant performance of the proposed GA-based approach compared to $k\text{-means}$ in all the evaluation parameters.
\par \textcolor{black}{In the future, instead of maximizing sum-rate, energy efficiency maximization can be considered. Further, obtaining a closed-form solution of the proposed objective can also be taken into account. Also, remaining resources at the UAVs, e.g., bandwidth, etc., can be assigned to cognitive radio users by considering a cognitive radio framework \cite{karamCRNWs, karamcrnws2} within our system model represented in Fig.\,\ref{systemmodel}. Furthermore, one can consider the hover time as an additional communication constraint \cite{uav3, new_R2}. To this end, signalling overhead of UAVs in the calculation of hover time/battery life etc., can be a potential future direction \cite{new_R6_signalling, new_R7_signalling}. Another important aspect could be to consider CSI acquisition in the UAVs network as it can help in improving the quality of precoding, etc. \cite{new_RNN}.}

%%%%%%%%%%%%%%%%%%%%%%%%%%%%%%%%%%%%%%%%%%%%%%%%%%%%%%%%%%%%%%%%%%%%%%%%%%%%%%%%%%%%%%%%%%%%%%%%%%%%%%%%%%%%%%%%%%%%%%%%%%%%%%%%%%%%%%%%%%%%%%%%%%%%%%%%%%%%%%%%%%%%%%%%%%%%%%%%%%%%%%%%%%%%%%%%%%%

\ifCLASSOPTIONcaptionsoff
  \newpage
\fi



%\begin{thebibliography}{1}

\bibliographystyle{IEEEtran}
%\bibliography{mybib}

%\end{thebibliography}
%\begin{biography}[{\includegraphics[width=1in,height=1.25in,clip,keepaspectratio]{mshell}}]{Michael Shell}

% ==== SWITCH OFF the BIO for submission
% ==== SWITCH OFF the BIO for submission
%\vskip -2\baselineskip plus -1fil
%
\begin{IEEEbiography}[{\includegraphics[width=1.3in,height=1.25in,clip,keepaspectratio]{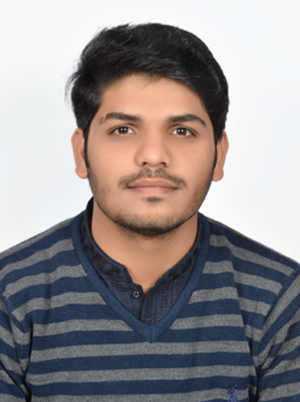}}]{Muhammad K. Shehzad} (S'21) received his M.S. in Electrical Engineering from NUST, Islamabad, Pakistan, in 2019. During his M.S., he also spent one semester on ERASMUS+ mobility program at University of Malaga (UMA), Malaga, Spain. He received his BEng. (Hons) in Electrical and Electronic Engineering from University of Bradford, Bradford, UK, in 2016. In addition, he holds student membership of IEEE. From 2016 to 2017, he worked as a Research Assistant at Namal Institute, Mianwali, Pakistan. During November 2019 to February 2020, he also served as Research Assistant at SEECS, NUST, Islamabad, Pakistan. Currently, he is working as a Research Engineer and Ph.D. student at Nokia Bell Labs and CentraleSupelec, Paris, France, respectively. His research interests include cognitive radio networks, UAV communication, databases, Internet of Things (IOTs), and multiple-input multiple-output (MIMO) communication using artificial intelligence (AI).
\end{IEEEbiography}
\vskip -2\baselineskip plus -1fil
\begin{IEEEbiography}[{\includegraphics[width=1in,height=1.25in,clip,keepaspectratio]{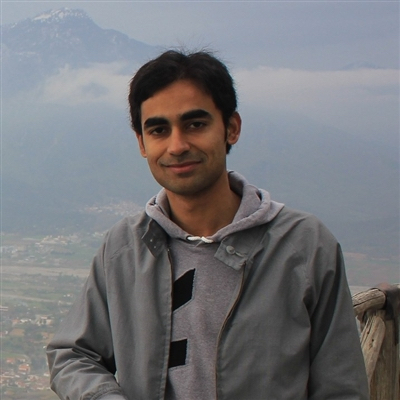}}]{Arsalan Ahmad} (S'12-M'14) received his Ph.D. in Electronics and Communication Engineering from Politecnico di Torino, Italy in Feb 2014. His research area was energy efficiency in optical network planning. He received his MS degree in Communication Engineering from Politecnico di Torino, Italy in 2010. After completing his Ph.D., he has worked as a post doc researcher at Politecnico di Torino, Italy on the physical layer aware design of fixed and flexible grid optical networks. He joined NUST in 2015 where he is currently working as an Assistant Professor and the director of Software Defined Optical Networks (SDON) lab. In the summer of 2019, he was a visiting research fellow at CONNECT research centre, Trinity College Dublin, Ireland. He serves as a reviewer of IEEE/OSA Journal of lightwave technology (JLT), Journal of optical communication and networking (JOCN) and Wiley International Journal of Communication Systems (IJCS). His main research interests include optical networks, physical layer aware optical networking and energy efficiency in network planning.
\end{IEEEbiography}
\vskip -2\baselineskip plus -1fil
\begin{IEEEbiography}[{\includegraphics[width=1in,height=1.25in,clip,keepaspectratio]{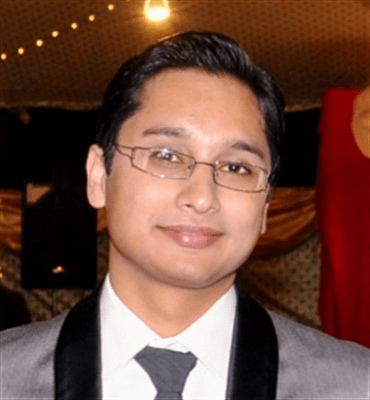}}]{Syed Ali Hassan}
(S'09-M'12-SM'17) received his Ph.D. in Electrical Engineering from Georgia Institute of Technology, Atlanta USA in 2011. He received his M.S. Mathematics from Georgia Tech in 2011 and M.S. Electrical Engineering from University of Stuttgart, Germany in 2007. He was awarded BE degree in Electrical Engineering from NUST, Pakistan, in 2004. His broader area of research is signal processing for communications with a focus on cooperative communications for wireless networks, stochastic modeling, estimation and detection theory, and smart grid communications. Currently, he is working as an Associate Professor at SEECS, NUST, where he is the director of Information Processing and Transmission (IPT) Lab, which focuses on various aspects of theoretical communications. He was a visiting professor at Georgia Tech in Fall 2017 and also holds senior membership of IEEE. He also held industry positions, in Cisco Systems Inc. CA, USA, and Center for Advanced Research in Engineering, Islamabad, Pakistan.
\end{IEEEbiography}
\vskip -2\baselineskip plus -1fil
\begin{IEEEbiography}[{\includegraphics[width=1in,height=1.25in,clip,keepaspectratio]{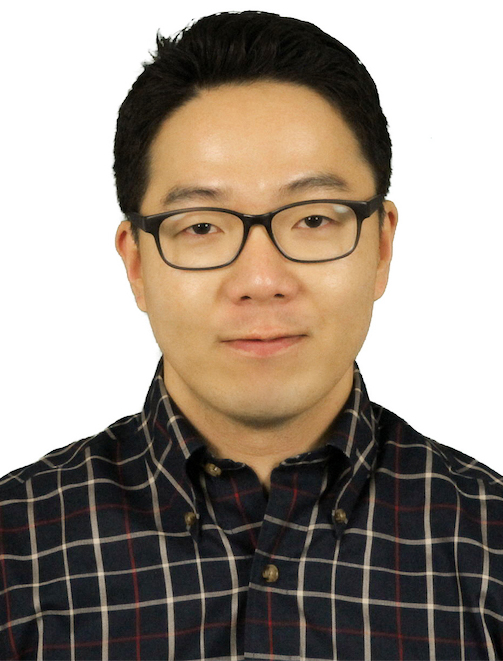}}]{Haejoon Jung} (S'10-M'14-SM'21)
received the B.S. degree (Hons.) from Yonsei University, South Korea, in 2008, and the M.S. and Ph.D. degrees from the Georgia Institute of Technology (Georgia Tech), Atlanta, GA, USA, in 2010 and 2014, respectively, all in electrical engineering. From 2014 to 2016, he was a Wireless Systems Engineer at Apple, Cupertino, CA, USA. He joined Incheon National University, Incheon, South Korea, in 2016, where he is currently an Associate Professor with the Department of Information and Telecommunication Engineering. His research interests include communication theory, wireless communications, wireless power transfer, and statistical signal processing.
 \end{IEEEbiography}

%\begin{IEEEbiographynophoto}{Ignacio Ramos}
%(S'12) received the B.S. degree in electrical engineering from the University of Illinois at Chicago in 2009, and is currently working toward the Ph.D. degree at the University of Colorado at Boulder. From 2009 to 2011, he was with the Power and Electronic Systems Department at Raytheon IDS, Sudbury, MA. His research interests include high-efficiency microwave power amplifiers, microwave DC/DC converters, radar systems, and wireless power transmission.

%% insert where needed to balance the two columns on the last page with
%%\newpage

% ==== SWITCH OFF the BIO for submission
% ==== SWITCH OFF the BIO for submission


\vfill


\end{document}